\documentclass[useAMS,usenatbib]{mn2e}
\usepackage[dvips]{graphicx}
\usepackage{t1enc}
\usepackage[latin1]{inputenc}
\usepackage[english]{babel}
\newcommand{\abs}[1]{\left|#1\right|}
\newcommand{\eps}{\varepsilon}
\newcommand{\vp}{\mathcal{P}}
\renewcommand{\exp}[1]{e^{#1}}
\newcommand{\ud}[1]{\mathrm{d}#1}
\newcommand{\br}[1]{\left(#1\right)}

\title[On the axisymmetric thin disk model]
{On the axisymmetric thin disk model of flattened galaxies}
\author[
{\L}ukasz Bratek, Joanna Ja{\l}ocha, Marek Kutschera]{ {\L}ukasz
Bratek$^{1}$, Joanna Ja{\l}ocha$^{1}$ and Marek
Kutschera$^{1,2}$\\ $^{1}$ H. Niewodnicza\'nski Institute of
Nuclear Physics,
Polish Academy of Sciences, Radzikowskiego 142, 31-342 Krak\'ow, Poland \\
$^{2}$ M. Smoluchowski Institute of Physics, Jagellonian
University, Reymonta 4, 30-059 Krak\'ow, Poland}
\begin{document}
\date{}
\pagerange{ \pageref{firstpage}--\pageref{lastpage}} \pubyear{}

\maketitle
\begin{abstract}
Non-monotonic features of rotation curves, and
also the related gravitational effects typical of thin disks -- like backward-reaction
or amplification of rotation by negative surface density gradients -- which
are characteristic imprints of disk-like mass distributions, are
discussed in the axisymmetric thin disk model.
The
influence of the data cutoff in rotational velocity measurements on the
determination of the mass distribution in flattened galaxies is studied.

It has also been found that the baryonic matter distribution in the spiral galaxy
NGC 5475, obtained in the axisymmetric thin disk approximation,
accounts for the rotation curve of the galaxy. To obtain these results,
 the iteration method developed recently by the
authors has been applied.

\end{abstract}


\section{Introduction}

The analysis of rotation curves provides the most reliable
 means for ascertaining at least the gross
distribution of gravitating matter within spiral galaxies, as Alar
Toomre pointed out \citep{bib:toomre_ad_disk_eq}. In the same paper Toomre
formulated a complete mathematical model of an axisymmetric and
infinitely thin disk rotating under its own gravity. The model
offers a tool for determining the equilibrium mass distribution
directly from the rotation law of a highly flattened system, such
as a spiral galaxy. It is assumed in the framework of the model,
that orbits of stars, gas, etc., are
circular, and that gravitational forces are balanced solely by
centrifugal forces; the effect of pressure is thus ignored. A few years earlier, following the work reported by
\citep{bib:prender}, Brandt discussed less general situation of a very flattened system regarded
as an assembly of osculating homoeoids compressed to a circular
disk \citep{bib:brandt}. All these papers followed
many other pioneering ones referred to in \citep{bib:brandt}.

The customary
parametric few-component  models relate the mass distribution of baryonic matter mainly to luminosity measurements. The obtained amount of luminous mass is usually insufficient to account for the observed rotation of galaxies and a massive spherical dark halo is introduced as a remedy for the missing mass.
The thin disk model with
the surface mass density reconstructed mainly from rotation curves performs very well with strongly
non-monotonic rotation curves, whereas the customary models have difficulties in explaining them. The thin disk model can easily account for high local gradients of rotational
velocity, typically giving
a lower total mass.
In an extreme example, the rotational velocity of an outer galactic region, falling off faster than Keplerian, cannot be explained by the presence of a massive spherically symmetric halo. However, such a feature may be explained using a flattened mass distribution with a suitably changing mass profile.

The major difficulty in the practical use of the disk model lies with unambiguous mass
density reconstruction. Unlike for spherical symmetry, the density depends on the assumed extrapolation of the rotation curve beyond the last measurement point (referred to in this paper as the
'cutoff radius'). This is a consequence of the nonlocal relation
between the surface density and the rotational velocity of a disk-like
system.
For a reliable reconstruction of the disk surface density,
the rotation curve has to be known globally (which, for observational reasons, is
impossible) or at least out
to its Keplerian falloff.\footnote{\cite{bib:22} report on galaxies with Keplerian
rotation curves at large radii} Therefore, the
velocity measurements must be supplemented with independent data
in order to constrain the sought mass distribution.
This may be achieved, for example, by taking into account the
amount of hydrogen measured in the outermost galactic regions as was done in \citep{bib:bratek}.

Mass estimates of spiral
galaxies are strongly model-dependent.
Rotational velocities, when interpreted in a
spherical dark halo model, may greatly overestimate galactic
masses as compared to the disk model. It is thus important to
consider a phenomenologically acceptable model that gives a
lower limit to the mass. The thin disk model may serve as such a reference
model.

NGC 4736 is an example of a spiral galaxy with a rotation curve of which outer parts cannot be satisfactorily reconstructed when a spherical halo is assumed.
By a simple examination one finds that in the circular orbit approximation, the rotation curve of this
galaxy cannot be created by a spherical matter distribution. By applying the thin disk model, with no constraining
assumptions about the mass-to-light ratio profile, one can find a
mass profile in this galaxy that perfectly conforms  with its rotation
curve, and agrees with the amount of hydrogen observed at large radii, beyond the
cutoff radius \citep{bib:bratek}. Only insignificant (if any) amount of dark matter is required, whereas the customary models predict the galaxy to be dark matter dominated.
In this paper we report also on another
spiral galaxy, NGC 5457, in which the amount of baryonic matter
found in the disk model accounts for this galaxy rotation.

\subsection{The global thin disk model as a means of relating the rotation law to
a mass distribution} The main objection against the use of the axisymmetric thin disk as a model of
a spiral galaxy, is its instability with too short time scale
\citep{bib:toomre_stab, bib:stab_numer}. The simplest way to stabilize such a system is to add a sufficiently strong spherically symmetric potential. This is usually done by introducing a massive spherical halo of dark matter, which is believed to surround spiral galaxies. However, it should be noted that the
galactic interior, irrespectively of its structure, produces at radii
sufficiently large almost spherically symmetric
potential that may stabilize the external
circular orbits.

Putting the stability issue aside, the disk model is still useful for approximate description
of the gravitational field of a flattened galaxy. Suppose that the rotation curve of such a galaxy represents the velocity of the streaming motion of matter rotating on
roughly circular orbits in the galactic disk. The
disk model associates with this rotation curve a formal
surface mass density that may be considered as an approximation of the column
density of matter in this galaxy. To ensure its applicability, we use
the model only for describing galaxies with rotation curves breaking the condition of spherical mass distribution at large distances from their centers. We assume that this non-sphericity can be attributed to flattening of these galaxies. In
these regions we therefore expect the disk model to better reflect the oblateness of such galaxies than the models with massive spherical halo.
What's more, the customary rotation curve modeling already assumes a
resultant thin disk surface density as a superposition of the
stellar disk component's surface density and of the column density of
the spherical bulge component projected onto the thin disk's
plane.
Based on these premises the use of the single global thin disk for modeling the whole mass distribution in a flattened galaxy seems justified.
The only new qualitative thing is that the  effective disk becomes extended
further out to the outer part in the case when the sphericity condition is broken
there, that is, when the massive spherical halo cannot be introduced.

\section{The sphericity problem}
The velocity of test bodies moving on circular orbits in
the equatorial plane of a spherically symmetric mass distribution with
a mass function $M(r)$ is given by $v(r)=\sqrt{GM(r)/r}$. By analogy
with this relation we define the Keplerian mass function $r\,v^2(r)/G$
of a galaxy with a rotation curve $v(r)$. For
a galaxy with a massive spherical CDM halo, the Keplerian mass function, at least
for large radii,
should agree  with the true mass function of the galaxy.

Keplerian mass functions of some spiral galaxies decline or are
non-monotonic in their outer parts, suggesting  the presence of an
extended, flattened and massive subsystem in these galaxies rather
than a spherical one. The spherical dark halo may be
excluded in favor of a large flattened subsystem if the sphericity
condition --
\begin{equation}\label{eq:sph_cond}\mathrm{for{\ }all{\ }}\,r:\,\br{r\,v^2(r)}'\geq0\end{equation}
-- is broken at larger radii (beyond the
central bulge). In this case the global disk model seems more appropriate for reconstruction
of the mass distribution.  Note,
that the central spherical bulge may be represented by a column
mass density in the disk plane since the details of the internal mass
distribution are not very important for the determination of the
gravitational field in the distant regions, thus the use of the disk model in the central regions is also acceptable.

\begin{figure*}
\begin{tabular}{|@{}c@{}|@{}c@{}|}
\includegraphics[width=0.333\textwidth]{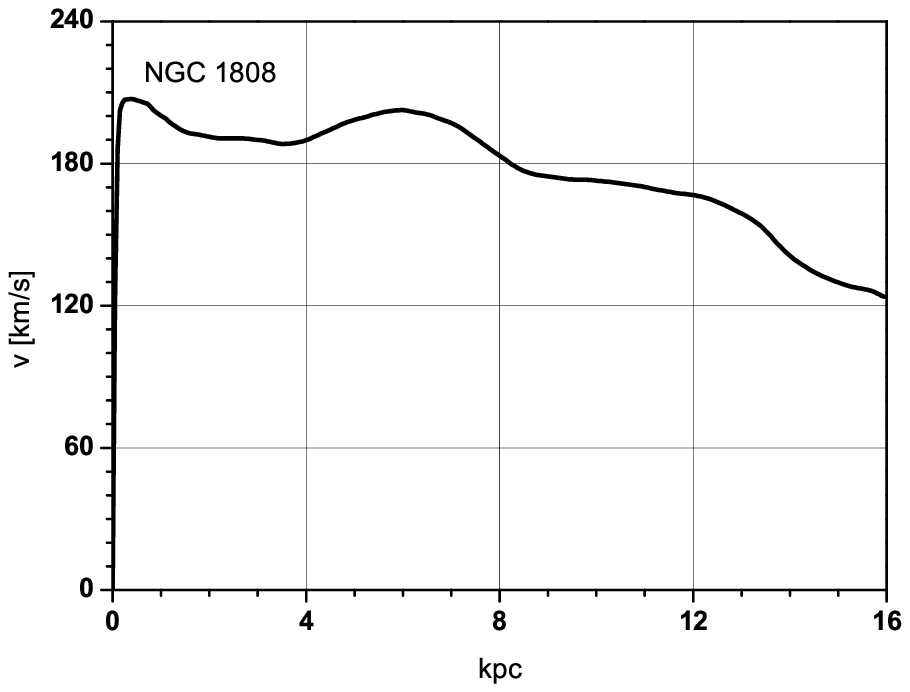}
&\includegraphics[width=0.333\textwidth]{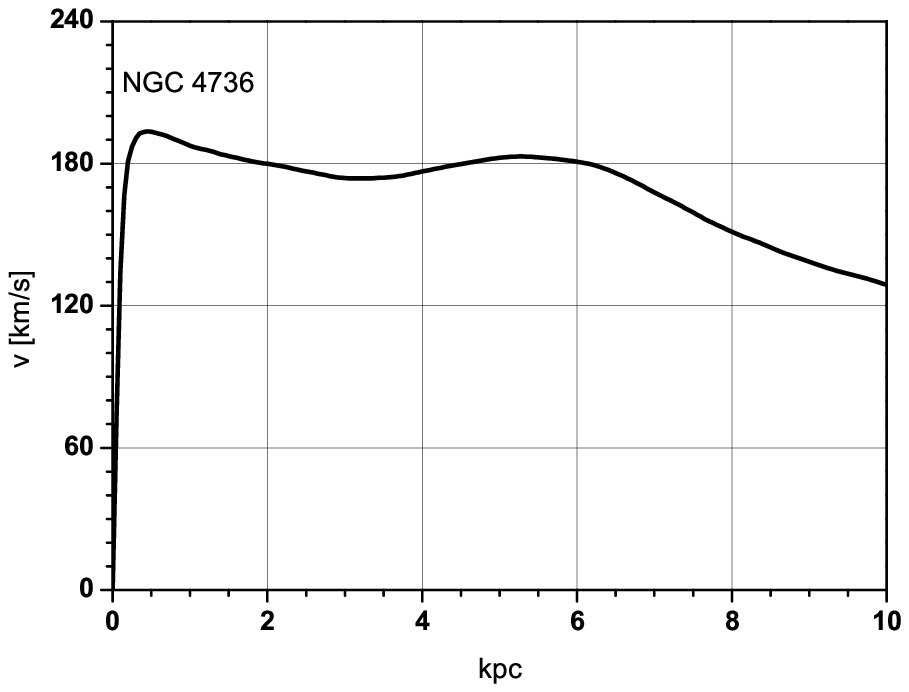}\\
\includegraphics[width=0.333\textwidth]{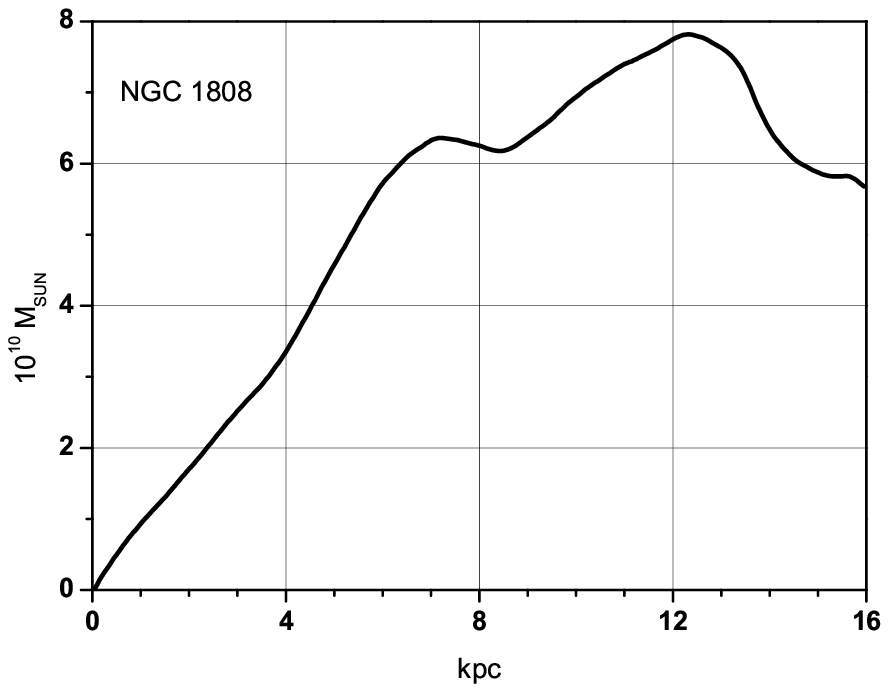}
&\includegraphics[width=0.333\textwidth]{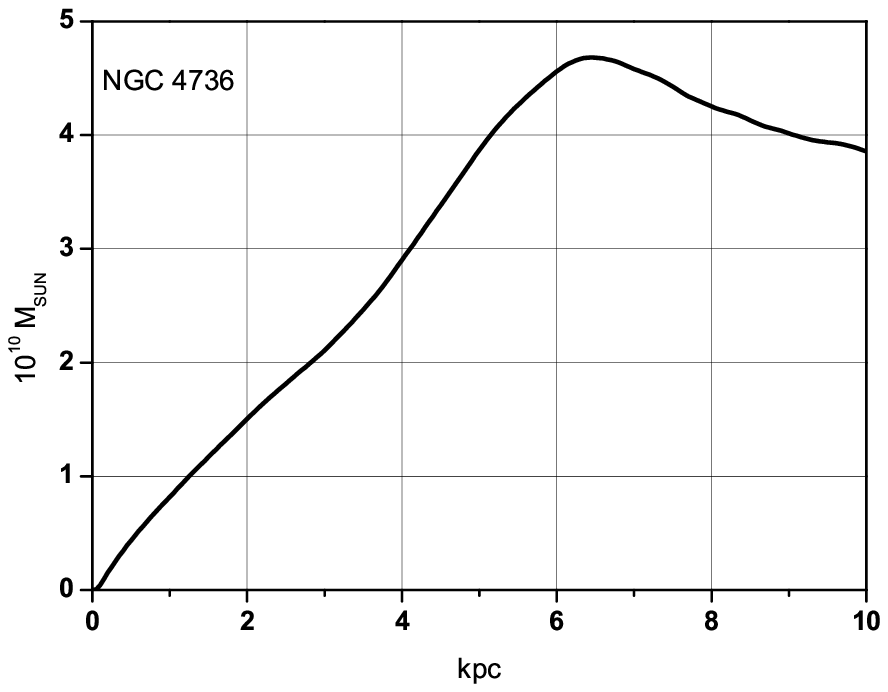}\\
a)&b)
\\\end{tabular} \caption{ \label{fig:test}
The upper plots show rotation curves of spiral galaxies
 \textbf{a)} NGC1808  and \textbf{b)} NGC4736. The lower plots 
show the corresponding Keplerian mass functions defined in the text.
The unusual behavior of these functions (a mass function should be nondecreasing) suggests that matter
distribution in these galaxies may be dominated by an
extended and flattened galactic subsystem.}
\end{figure*}

Since there are many examples of disk-like mass configurations 
with their rotation laws satisfying the sphericity condition
(\ref{eq:sph_cond}), violation of the condition at larger radii
provides a strong argument for the presence of an extended flattened component
and against a significant spherical halo. For example, the Toomre thin disk of unit
mass   with the surface mass density
$\sigma(x)=(1+x^2)^{-3/2}/\br{2\pi}$ ($x$ is a dimensionless
radius in cylindrical coordinates), rotates differentially with
velocity $v(x)=x\br{1+x^2}^{-3/4}$ \citep{bib:toomre_ad_disk_eq},
hence $\br{xv^2(x)}'= 3x^2\br{1+x^2}^{-5/2}>0$. As so, a spherically symmetric mass
distribution is possible with the same rotation curve. In the approximation of circular orbits, the
sphericity condition is thus necessary, although not sufficient,
for the presence of an extended and dominating spherical halo. Figure \ref{fig:test} shows two rotation curves that
break inequality (\ref{eq:sph_cond}) at large radii.

\section{Rotation in the gravitational field of an axisymmetric flat disk}
\subsection{Basic equations}
Rotational velocity $v(\rho)$  of circular motion of a test body
around the symmetry axis and in the symmetry plane of the
gravitational field of an axisymmetric thin disk  is related to the
disk's surface mass density $\sigma(\rho)$ by
\begin{eqnarray}\label{eq:vfromsig}
\frac{v^2(\rho)}{4\,G\,\rho}=\vp\left[
\int\limits_{0}^{\rho}\sigma(\chi)\frac{\chi{}E
\br{\frac{\chi}{\rho}}}{\rho^2-\chi^2}\,\ud{\chi}\right. \nonumber\\
\left.-\int\limits_{\rho}^{\infty}\sigma(\chi)\biggl(
\frac{\chi^2E\br{\frac{\rho}{\chi}}}{\rho\br{\chi^2-\rho^2}} -
\frac{K\br{\frac{\rho}{\chi}}}{\rho}\biggr)\ud{\chi}\right]
,\end{eqnarray} see appendix  \ref{app:deriv}.\footnote{$\rho$ is the radial variable in cylindrical
coordinates and $K$ and $E$ are complete elliptic functions of the
first and second kind, \citep{bib:15}\[
K(\kappa)=\int\limits_{0}^{\pi/2}\frac{\ud{\phi}}{
\sqrt{1-\kappa^2\sin^2{\phi}}},\qquad
E(\kappa)=\int\limits_{0}^{\pi/2}\ud{\phi}
\sqrt{1-\kappa^2\sin^2{\phi}}\]} The above integral
exists in the 'principal value' sense, hence the $\vp$ symbol. The
second part of the integral describes the backward interaction
effect typical for disks  --
 the velocity of rotation on a circular orbit is influenced also by matter present outside
the orbit. Gravitation of disks is therefore qualitatively
different from that of spherically symmetric systems. Note,
that for arbitrary $\sigma>0$, integral (\ref{eq:vfromsig}) may
attain negative values at some radii. For such radii circular
orbits are impossible.

The formal inverse of
(\ref{eq:vfromsig}) reads
\begin{eqnarray}\label{eq:sigmamoja}\sigma(\rho)=
\frac{1}{\pi^2G} \vp \left[\int\limits_0^\rho
v^2(\chi)\biggl(\frac{K\br{\frac{\chi}{\rho}}}{ \rho\
\chi}-\frac{\rho}{\chi}
\frac{E\br{\frac{\chi}{\rho}}}{\rho^2-\chi^2}\biggr)\ud{\chi}
\right.\nonumber \\ + \left. \int\limits_\rho^{\infty}v^2(\chi)
\frac{E\br{\frac{\rho}{\chi}}
}{\chi^2-\rho^2}\,\ud{\chi}\right].\end{eqnarray}
 This integral exists if $\lim\limits_{\rho\to0}\rho\, v(\rho)=0$, however in its derivation more stringent
assumptions were imposed on $v(\rho)$ (see appendix
\ref{app:deriv}).

In the approximation of cold, circular
streaming motion, relation (\ref{eq:sigmamoja}) may be regarded as the definition of an effective surface mass density corresponding to the rotation curve of a flattened
disk-like galaxy. However, the answer to the question how accurately
the resulting gravitational potential approximates the true one in this galaxy,  would require separate studies, which are outside the scope of this paper. Nevertheless, it is still interesting to find out whether the effective surface density, that by construction accounts for the galaxy rotation, would also be consistent with other observational data available for the galaxy.   Such a procedure led to consistent results for the galaxy NGC 4736 in \citep{bib:bratek} and for the galaxy NGC 5457 in this paper.

Another relation
between the rotation velocity and the corresponding surface density is \citep{bib:toomre_ad_disk_eq}
\begin{eqnarray}\label{eq:toomre}
\widetilde{\sigma}(\rho)=\frac{1}{\pi^2G}\ \vp \left[
\int\limits_0^{\rho}\frac{\ud{v^2}(\chi)}{\ud{\chi}}
\cdot\frac{1}{\rho}K\br{\frac{\chi}{\rho}}\ud{\chi}\right.\nonumber
\\+\left.\int\limits_{\rho}^{\infty}\frac{\ud{v^2}(\chi)}{\ud{\chi}}
\cdot\frac{1}{\chi}K\br{\frac{\rho}{\chi}}\ud{\chi}\right],
\end{eqnarray} where again $\vp$ denotes a principal value integral.
This formula became popular mainly thanks to the handbook by
\cite{bib:binney}. Provided
$\lim_{\rho\to0}v^2(\rho)\to0$, formulas (\ref{eq:sigmamoja}) and
(\ref{eq:toomre}) are equivalent
$\sigma(\rho)\equiv\widetilde{\sigma}(\rho)$ unless integration
gets cut off at a finite radius. However, the disadvantage in the practical use of the formula (\ref{eq:toomre}) is that it contains the
first derivative of $v^2(\rho)$ which is subject to high
observational errors that substantially enlarge the uncertainties in
the determination of the surface density.

\subsubsection{\label{sec:example}An example}
To see  how the basic equations of disk model work, we may apply them to the most familiar example provided by the Mestel disk.\footnote{The example is simple but unfortunate since Mestel's disk is incompatible with the mathematical consistency of the disk model, which was referred to in the appendix \ref{app:deriv}, however, by doing calculations carefully, it can be still considered.}
Let  $\sigma(\rho)=\frac{M}{2\pi\,a\,\rho}$ be the disk surface density. On substituting
$\chi=x\rho$ and $\chi=x^{-1}\rho$, respectively, in the first and
in the second integral in (\ref{eq:vfromsig}), we obtain
$$\frac{v^2(\rho)}{4G\,\rho}=\frac{M}{2\pi\,a\,\rho}\times
\left\{\int\limits_{0}^{1}
\frac{\br{1+x}K(x)-E(x)}{x(1+x)}\ \ud{x}=\frac{\pi}{2}\right\},$$ that is,
the rotational velocity is constant: $v^2(\rho)=\frac{G\,M}{a}$. Reversely,
for a constant velocity $v_o=\frac{G\,M}{a}$, we
obtain from (\ref{eq:sigmamoja}), by similar substitutions as
before,
$$\sigma\br{\rho}=\frac{v_o^2}{\pi^2G\rho}\int\limits_{0}^{1}
\frac{\br{1+x}K(x)-E(x)}{x(1+x)}\
\ud{x}=\frac{v_o^2}{2\pi\,G\rho}=\frac{M}{2\pi a \rho}.$$
The textbook equation
(\ref{eq:toomre}) cannot be used  for Mestel disk directly, since
 its rotation curve violates the requirement $\lim\limits_{\rho\to0}v(\rho)=0$ (otherwise, one would conclude that
  a disk rotating with constant velocity has vanishing surface density).  However, if we write
$v^2(\rho)=v_o^2\Theta\br{\rho-\eps}$ with small $\eps>0$, where
$\Theta$ stands for the unit step function, then
$v^2{'}(\rho)=v_o^2\delta(\rho-\eps)$, and now we obtain from (\ref{eq:toomre}) $\sigma(\rho)=\frac{v_o^2K(\eps/\rho)}{\pi^2G\,\rho}$. In the
limit $\eps\to0$ we obtain the correct result
$\sigma(\rho)=\frac{v_o^2}{2\pi\,G\,\rho}$ for $\rho>0$, since
$K(0)=\frac{\pi}{2}$. Again the formula (\ref{eq:sigmamoja}) has proved more advantageous than the textbook formula (\ref{eq:toomre}).

\subsection{Some consequences of the nonlocal
correspondence between $\sigma$ and $v$} 
\begin{figure*}
\includegraphics[width=0.333\textwidth]{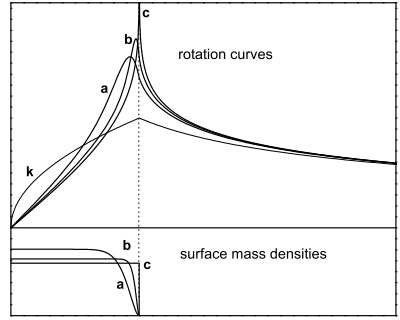}
\caption{\label{fig:disksmpl} The solid lines marked by \textbf{a}, \textbf{b}, and \textbf{c}, show respectively the elements $n=16$, $n=50$, and the
limit $n\to\infty$ of the sequence of unit mass disks with the surface
mass densities $\sigma_n$ defined in equation (\ref{eq:sequence}),
and the corresponding rotational velocities. The rotational velocity in the equatorial plane of a ball with  the same mass function as that of the limiting disk is marked by \textbf{k}.}
\end{figure*}

Rotational velocities of disks may
attain high values in
the regions where the surface mass density decreases rapidly, and  can be much larger than rotational velocities of spherically symmetric matter
configurations of the same internal mass. To illustrate this, let us consider a sequence of unit mass disks with the
surface mass densities
\begin{eqnarray}\label{eq:sequence}\sigma_{n}(x)=
\frac{3+n}{2\pi(n-1)}\left(2-(n+1)x^{n-1}\right.\nonumber\\
+\left.(n-1)x^{n+1}\right)\Theta(1-x),\qquad n>2.
\end{eqnarray} $\Theta$ is the unit step function and $x$ is a dimensionless radial
variable. For $n$ sufficiently large, $\sigma_{n}(x)$'s
are almost constant and fall off rapidly to $0$ close to $x=1$.
Figure \ref{fig:disksmpl}
illustrates the surface densities $\sigma_{n}(x)$ and the
corresponding rotational velocities $v_n(x)$. The functional
sequence $v_n(x)$ attains the theoretical limit
\[v_{\infty}(x)=\left\{\begin{array}{cc}
\sqrt{\frac{2}{\pi}\br{K(x)-E(x)}},& 0<x<1\\
\sqrt{\frac{2}{\pi}x\br{K(x^{-1})-E(x^{-1})}},& x>1
\end{array}\right.,\] which is unbounded at $x=1$. This signalizes
that for large $n$  and close to $x=1$, the rotational velocity can be
much larger than the Keplerian velocity $v_K(x)=\sqrt{M(x)/x}$, where $M(x)$ is the mass function.
This simple example illustrates that outside a highly oblate
object, where densities are already very small or negligible,
the rotation curve may be still non-Keplerian even at distances
from the object comparable with its size.

\begin{figure*}
\begin{tabular}{@{}c@{}c@{}c@{}}
\includegraphics[width=0.333\textwidth]{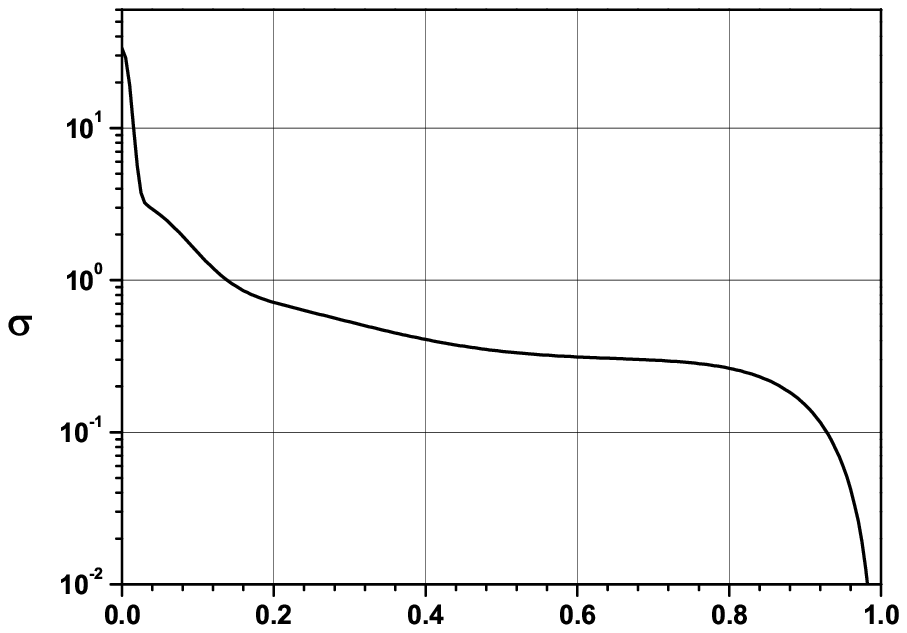}
&\includegraphics[width=0.333\textwidth]{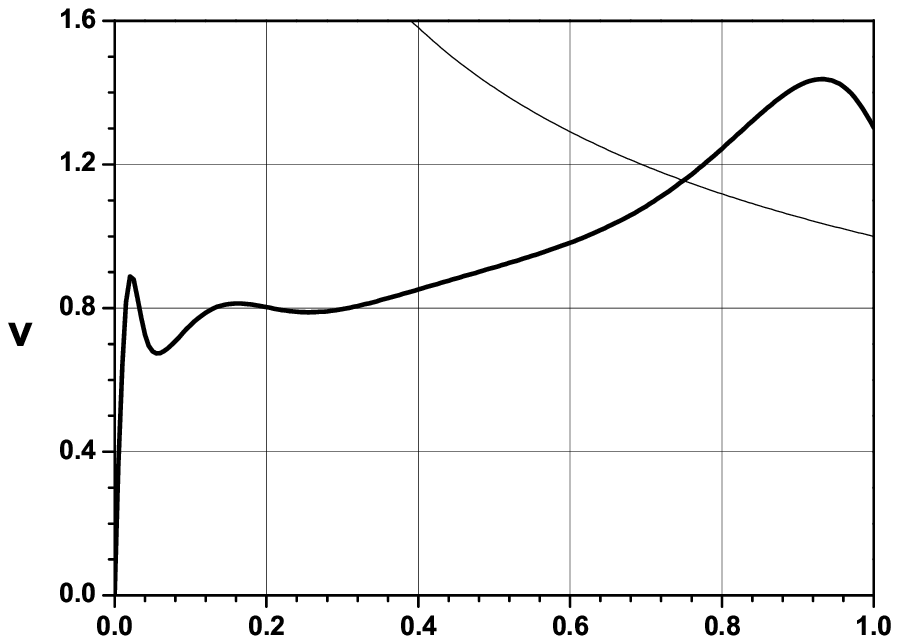}
&\includegraphics[width=0.333\textwidth]{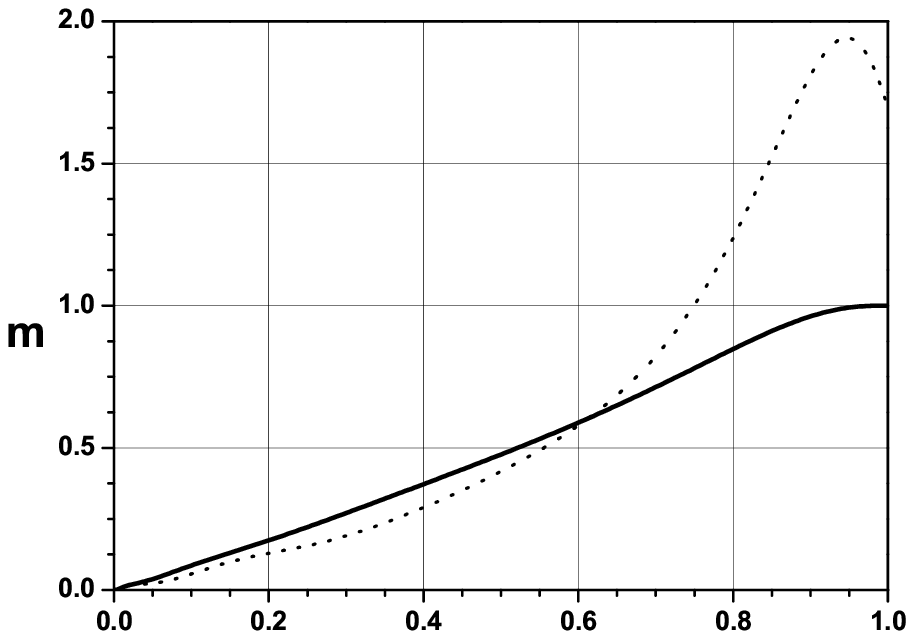} \\
a)&b)&c)\end{tabular} \caption{\label{fig:model} \textbf{a)}
The surface mass density
 of a unit mass disk: $\sigma(x)=\frac{100}{51\pi}\br{48\exp{-6400x^2}+
4\exp{-400x^2/3}+\exp{-100x^2/9}}+\frac{19}{40\pi}
\br{2-17x^{15}+15x^{17}}\Theta(1-x)$.
\textbf{b)} The corresponding rotation curve (thick line) and its Keplerian
asymptote (thin line). \textbf{c)} The true mass function (solid line) and the
Keplerian mass function calculated from the rotation curve (dotted
line). Although there is only $3\times10^{-6}$ of the total mass
outside the radius $x=1$, the rotation curve is still non-Keplerian
and the Keplerian mass is almost twice as big as the true mass. To explain this rotation curve at large radii, one would normally add a massive spherical halo by fitting it into the rising part with the help of the least square method. In this way additional 'missing' mass would be introduced to the system to account for its rotation, while in fact this mass is not present at all.
}
\end{figure*}

Figure \ref{fig:model} shows another example of a differentially rotating disk. Its rotation curve resembles qualitatively rotation curves of some spiral galaxies.
The first maximum in the rotation curve in figure \ref{fig:model}(b) reflects the contribution from a central bulge,  while the second maximum -- from an exponential disk. The outermost data points on this curve would influence mainly the dark matter halo fit. If
this model were to represent a real galaxy, the rotation curve could be measured
only out to $x\approx1$, since beyond this radius
only $3\times10^{-6}$ of the total mass would be present, thus amount hardly detectable. In effect,
the total estimated mass predicted by the customary models with
dark halos, would be roughly equal to the Keplerian mass evaluated
at the cutoff radius -- the true mass
in this example would be therefore overestimated by a factor of $\approx1.7$.

\begin{figure*}
\begin{tabular}{@{}c@{}c@{}c@{}}
\includegraphics[width=0.333\textwidth]{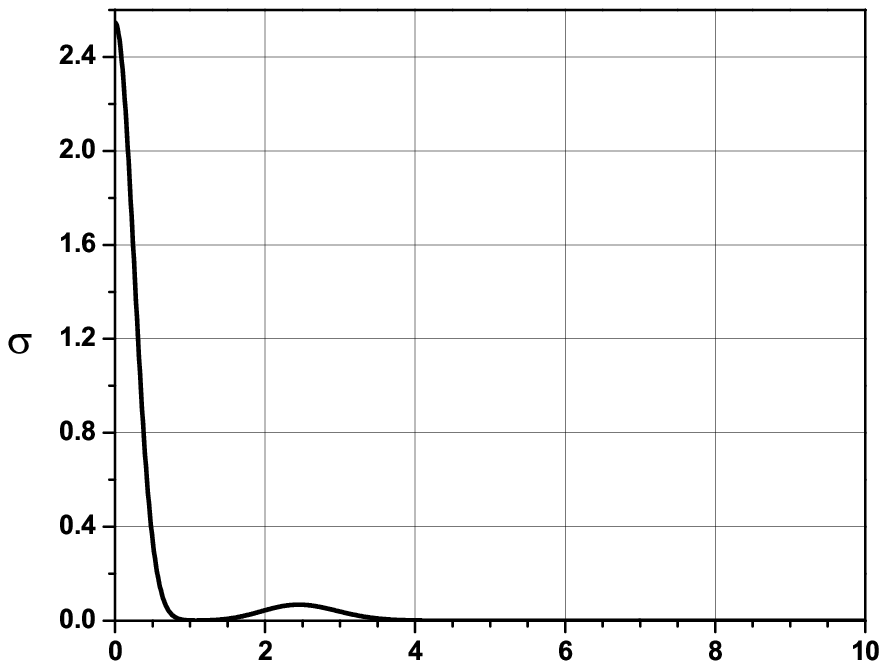}
&\includegraphics[width=0.333\textwidth]{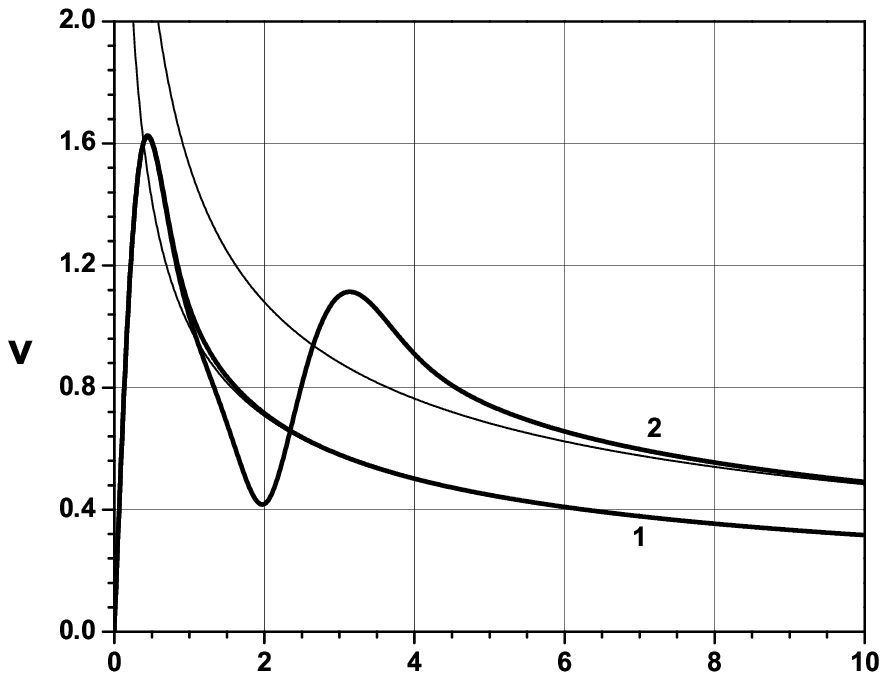}
&\includegraphics[width=0.333\textwidth]{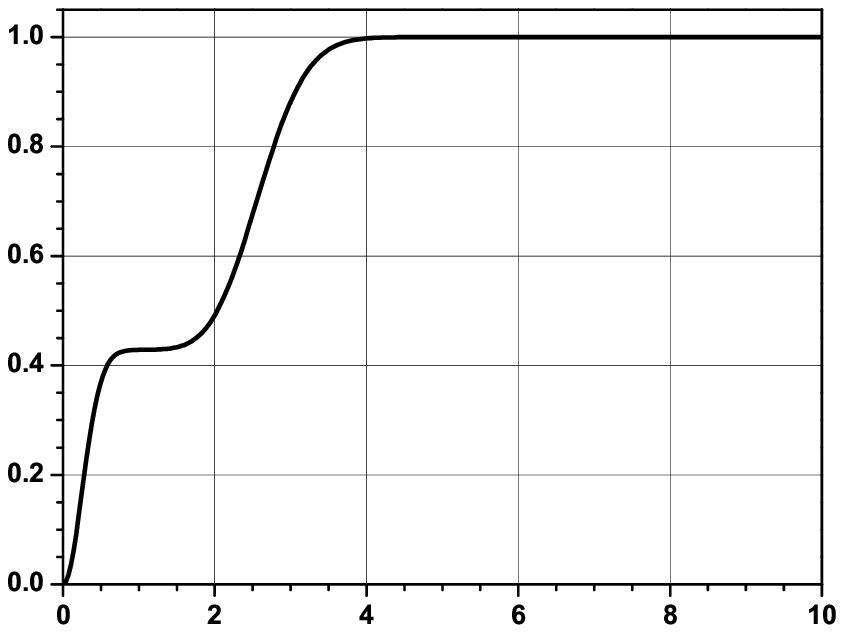}\\ a)&b)&c)
\end{tabular} \caption{\label{fig:ring} The influence of a massive external ring on the value of rotational velocities of test bodies moving on the internal orbits. \textbf{a)} The surface mass density
of an axisymmetric system composed of an internal disk coplanar with an external ring; \textbf{b)} the
corresponding rotation curves (and their Keplerian asymptotes) in the situation with (2) and without (1) the external ring; \textbf{c)} the mass function. Note, that in the presence of the external ring the rotational velocities in the vicinity of the internal disk are lower than those in the situation without the ring. }
\end{figure*}

It is a well known fact, that external spherically symmetric
shells of matter do not influence the motions of stars on internal
orbits, this is also true when the shells of constant density are
homoeoids \citep{bib:binney}. In general, however, the motions of stars are influenced by the presence of external masses. This backward reaction effect
is particularly important for disk-like objects. To illustrate this,
consider an axisymmetric system composed of a finite disk coplanar with an external massive ring. Figure \ref{fig:ring}
shows the rotational velocity of test bodies around the center of the system. In the presence of the external ring the velocities are lower than those estimated based on the internal masses only. In the case when the external ring is sufficiently close
or massive, such orbits may be impossible (the same is true if the internal mass is spherically symmetric). The presence of external flattened shells of matter may help to explain the very declined parts of rotation curves of some galaxies. It is evident from figure \ref{fig:ring} that the rotation of a very oblate
system can change much with radius depending on the density
gradients. In general, rotational velocities are amplified by
negative and attenuated by positive gradients of the disk surface
density. The gradients are not less important for the local value of the rotational velocity than the
amount of matter.

The elementary facts of disk gravitation discussed above,
indicate that mass models of flattened galaxies should be
sufficiently general in order to take the disk specific effects into
account. An important information about the mass distribution encoded in
the rotation curve may be simply overlooked
 by fitting a too much simplified parametric model, or by assuming
 the mass density in a luminous galactic subsystem to be proportional to its brightness profile. This in effect may lead to completely different
qualitative and quantitative  predictions for the same galaxy.  A typical example of
this situation is provided by the already mentioned galaxy NGC 4736.

\section{The cutoff error in the disk model}\label{sec:cutofef}

As follows from equation (\ref{eq:sigmamoja}), the mass function of a disk-like system is
$$M(\rho)=\frac{2}{\pi\,G}\int\limits_0^{\rho}\ud{\tilde{\rho}}\br{\vp
\int\limits_0^{\infty}
v^2\br{\tilde{\rho}
\,x}\mathcal{H}(x)\,\ud{x}}$$
where
$$\mathcal{H}(x)=\frac{\Theta\br{1-x}}{x}\br{K(x)-\frac{E(x)}{1-x^2}}+
\Theta\br{x-1}
\frac{E\br{x^{-1}}}{x^2-1}.$$
The total mass contained within radius $\rho$ can be therefore determined only when
a rotation curve is known globally, whereas the rotation curve of a real galaxy is known only out to a finite distance from its center (we call the distance the cutoff radius). This is the mathematical reason why the surface mass density and the mass function of a flattened galaxy cannot be determined uniquely from the measurements of its rotation curve only. This situation is drastically different from that for a spherically symmetric system in which case the mass function at a given radius is related directly to the local value of rotational velocity: $M(r)=r\,v^2(r)/G$.

Let $R$ denote the cutoff radius. The integration in (\ref{eq:sigmamoja}) gets naturally cut off at $\rho=R$, and for $\rho<R$ we may approximate (\ref{eq:sigmamoja})  by
\begin{eqnarray}\label{eq:sigmar}
\sigma_R(\rho)=\frac{1}{G\pi^2} \vp\left[\int\limits_0^\rho
v^2(\chi)\br{\frac{K\br{\frac{\chi}{\rho}}}{ \rho\
\chi}-\frac{\rho}{\chi}
\frac{E\br{\frac{\chi}{\rho}}}{\rho^2-\chi^2}}\ud{\chi}\right.
\nonumber\\ +\left.\int\limits_\rho^{R}v^2(\chi)
\frac{E\br{\frac{\rho}{\chi}} }{\chi^2-\rho^2}\ud{\chi} \right],
\qquad \rho<R.
\end{eqnarray}
Since $\sigma(\rho)\geq\sigma_R(\rho)$, expression
(\ref{eq:sigmar})  gives the surface density underestimated with
respect to (\ref{eq:sigmamoja}) by a cutoff error
 $\Delta\sigma_R(\rho)=\sigma(\rho)-\sigma_R(\rho)$:
 \begin{equation}
\Delta\sigma_R(\rho)=\frac{1}{G\pi^2}
\int\limits_{R}^{\infty}v^2(\chi) \frac{E\br{\frac{\rho}{\chi}}
}{\chi^2-\rho^2}\ud{\chi}>0,\qquad \rho<R.
\end{equation}
This error is completely unknown.
With the Toomre integral (\ref{eq:toomre}) cut off
at $\rho=R$ the situation is even worse. In this case we have for
$\rho<R$
\begin{eqnarray}\label{eq:sigRbar}
\widetilde{\sigma}_R(\rho)=\frac{1}{\pi^2G}\ \vp\left( \int
\limits_0^{\rho} \frac{\ud{v^2}(\chi)}{\ud{\chi}} \cdot
\frac{1}{\rho} K\br{\frac{\chi}{\rho}}\ud{\chi}\right.
\nonumber\\+\left.\int\limits_{\rho}^{R}
\frac{\ud{v^2}(\chi)}{\ud{\chi}} \cdot\frac{1}{\chi}
K\br{\frac{\rho}{\chi}}\ud{\chi}\right ),\qquad \rho<R,
\end{eqnarray} thus even the sign of the resulting cutoff error
$\Delta\widetilde{\sigma}_R(\rho)=\widetilde{\sigma}(\rho)-
\widetilde{\sigma}_R(\rho)$ is unknown, where
\begin{equation} \label{eq:deltasigRbar} \Delta\widetilde{\sigma}_R(\rho)=\frac{1}{G\pi^2}
\int\limits_{R}^{\infty} \frac{\ud{v^2}(\chi)}{\ud{\chi}}
\cdot\frac{1}{\chi} K\br{\frac{\rho}{\chi}}\ud{\chi},\qquad
\rho<R,
\end{equation} since it depends on a weighted slope of
the unknown part of the rotation curve.
It should be stressed that although
$\widetilde{\sigma}(\rho)\equiv{\sigma}(\rho)$, the corresponding
cut off integrals are different:
$\sigma_R(\rho)\ne\widetilde{\sigma}_R(\rho)$, thus also different
are the cutoff errors,
$\Delta\sigma_R(\rho)\ne\Delta\widetilde{\sigma}_R(\rho)$.

A cutoff error
of mass determination of the internal disk of radius $R$, and corresponding to $\Delta\widetilde{\sigma}_R$, reads
\[\Delta\widetilde{M}_R=\frac{Rv_R^2}{G}\cdot\frac{2}{\pi}
\int\limits_0^1x\ud{x}\int\limits_1^{\infty}
\frac{\ud{\xi}}{\xi}\frac{\ud{}u^2(\xi)}{\ud{\xi}}
K\br{\frac{x}{\xi}},\]  where $x=\rho/R$, $u(x)=v(x\,R)/v_R$ and
$v_R=v(R)$.

Let us consider a theoretical situation of a disk-like galaxy with
a rotation curve known to be almost Keplerian outside a cutoff radius
$R$. One then expects all of the galaxy mass $M$ to be contained
in the disk of radius $R$. Then also $M\approx{}G^{-1}R\,v^2_R$
and $u(x)\approx1/\sqrt{x}$ for $x>1$. By using
$\widetilde{\sigma}_R(\rho)$ to estimate the mass distribution in this
particular situation, we obtain for $0<x<1$
\begin{equation}\label{eq:deltasigma}\frac{1}{4\pi}<\frac{RG}{v_R^2}\cdot
\abs{\Delta\widetilde{\sigma}_R(\rho)}=
\frac{E(x)-(1-x^2)K(x)}{\pi^2x^2}< \frac{1}{\pi^2},
\end{equation}
thus the resulting total mass is greater by
$|\Delta\widetilde{M}_R|=(4/\pi-1)M\sim0.27M$ than the actual mass
$M$. The cutoff error $\abs{\Delta\widetilde{\sigma}_R(\rho)}$ in
(\ref{eq:deltasigma}) grows with the distance from the galactic
center.

This example leads to the following criterion:
provided that most of the disk mas is contained inside a radius
$R$, the cutoff error is insignificant in the region where
$\widetilde{\sigma}_R(\rho)$ satisfies the inequality
\begin{equation}\label{eq:criterion}\widetilde{\sigma}_R(\rho)>
\frac{1}{\pi^2}\frac{v_R^2}{RG}.
\end{equation}
Otherwise, the approximation (\ref{eq:sigRbar}) is not reliable, and
this occurs at larger radii, close to $R$. In the same theoretical
situation, the cutoff error corresponding to the integral
(\ref{eq:sigmar}), is given for $0<x<1$ by
\begin{equation}\frac{RG}{v_R^2}\cdot
\Delta\sigma_R(\rho)=\frac{1}{\pi^2}
\int\limits_1^{\infty}\frac{\ud{\xi}}{\xi}
\frac{E\br{\frac{x}{\xi}}}{\xi^2-x^2},\end{equation} hence
\[\frac{1}{4\pi}<-\frac{E(x)\ln(1-x^2)}{2\pi^2x^2}<\frac{RG}{v_R^2}\cdot
\Delta\sigma_R(\rho)<-\frac{ \ln(1-x^2)}{4\pi\,x^2},\] where the
inequalities $1=E(1)<E(x)<E(x/\xi)<E(0)=\pi/2$ for $x\in(0,1)$ and
$\xi>1$ have been used. The rightmost expression is divergent as
$x\to1$, but for $x<x_c\approx0.63$ it is still less than
$\pi^{-2}$, thus $\Delta\sigma_R(\rho)$ and
$\abs{\Delta\widetilde{\sigma}_R(\rho)}$ are comparable for
$\rho<0.63R$. In general, however, nothing is known about the mass
distribution beyond $R$, therefore the cutoff error can not be
even roughly estimated -- for if, e.g., outside a cutoff radius $R$ there existed a ring
of dark matter concentric with the disk, like in figure
\ref{fig:ring}, then the error could be arbitrarily large
depending on the undetectable annulus' mass.

\begin{figure*}
\begin{tabular}{@{}c@{}c@{}c@{}}
\includegraphics[width=0.333\textwidth,height=0.333\textwidth]{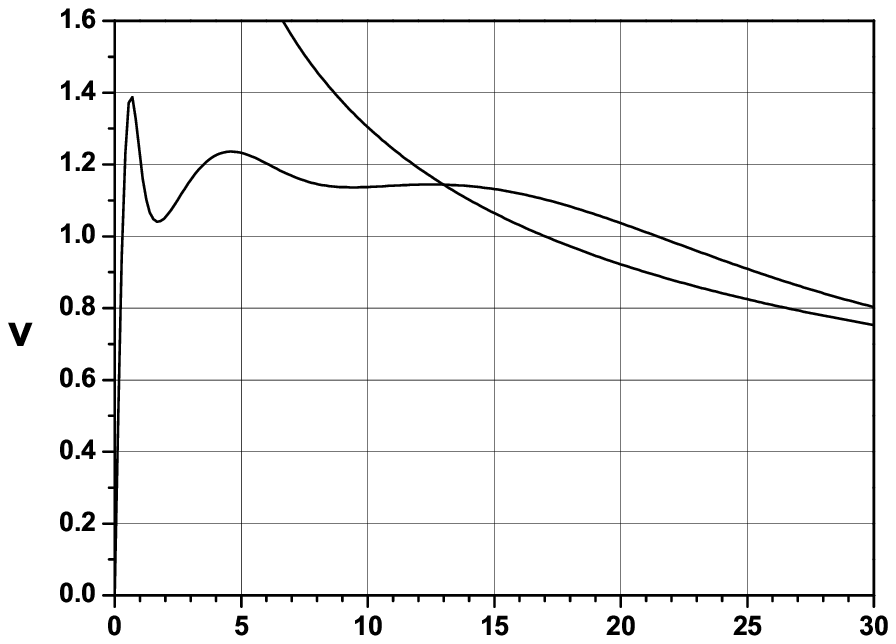}
&\includegraphics[width=0.333\textwidth,height=0.333\textwidth]{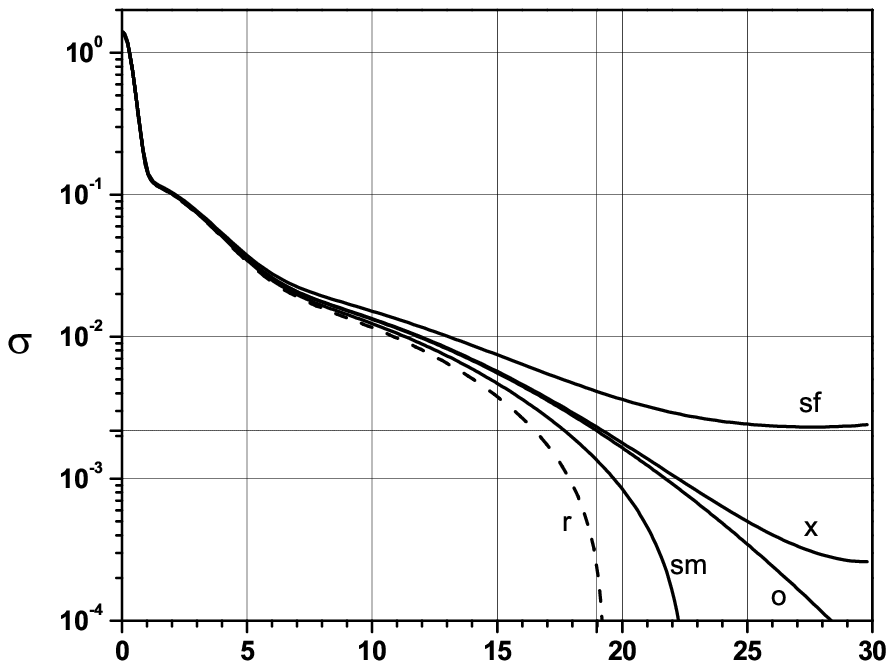}&
\includegraphics[width=0.333\textwidth,height=0.333\textwidth]{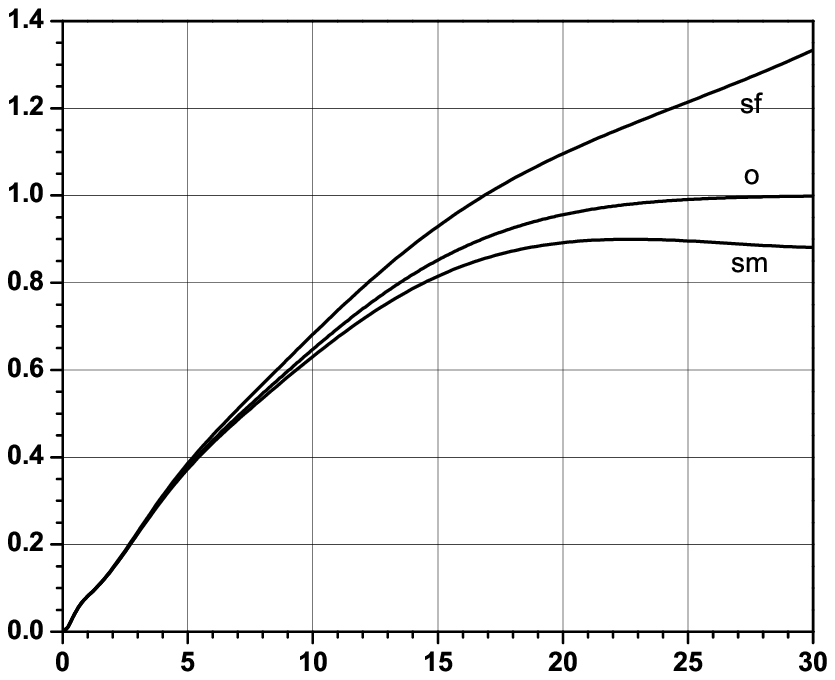}\\
a)&b)&c)\end{tabular} \caption{\label{fig:trisigma30}{\bf a)} The rotation curve (and
its Keplerian asymptote) of a disk
with the surface mass density
$\sigma(\rho)=\frac{4}{\pi}\br{\exp{-4\rho^2}+
\frac{1}{12}\exp{-\rho^2/12}+ \frac{1}{48}\exp{-\rho^2/144}}$. {\bf b)} The surface mass density reconstructed with the help of different methods
from the disk's rotation curve cut off at $R=30$: $o$ -- the exact profile, $sf$ -- the one calculated
from equation (\ref{eq:sigRbar}), $r$ -- from equation
(\ref{eq:sigmar}), $x$ -- the $sf$ surface density corrected for
the contribution from the Keplerian tail according to equation
(\ref{eq:deltasigma}), $sm$ -- the surface density obtained from a spectral representation of the
rotation curve (the internal mass was underestimated by $12\%$) -- the surface mass
density was calculated in this case from equation
(\ref{eq:sigmaseries}). The horizontal line represents the critical
value of the surface mass density defined in equation
(\ref{eq:criterion}), the abscissa of the vertical line is
$\approx0.63R$, which is the radius, to which $sf$ and $r$ may be
considered comparable. {\bf c)} The corresponding mass functions. }
\end{figure*}

\begin{figure*}
\begin{tabular}{|@{}c@{}|@{}c@{}|@{}c@{}|}
\includegraphics[width=0.333\textwidth,height=0.333\textwidth]{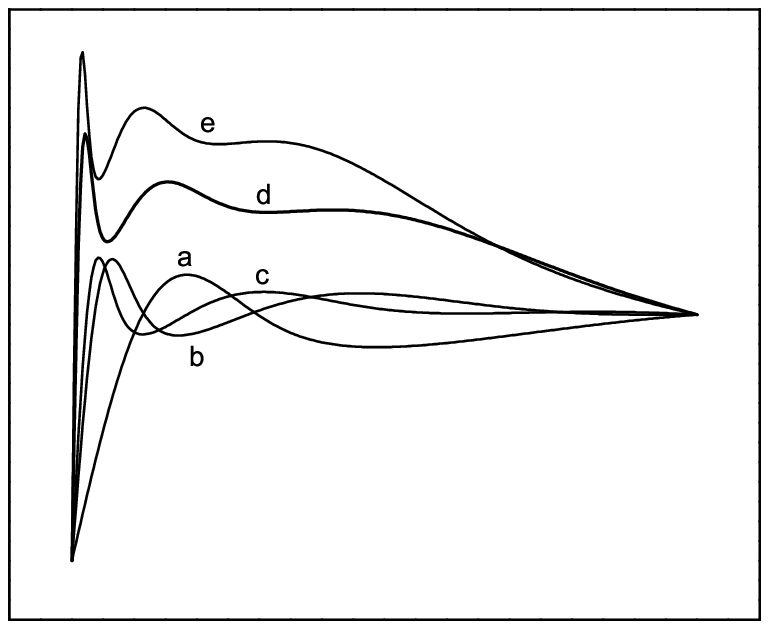}&
\includegraphics[width=0.333\textwidth,height=0.333\textwidth]{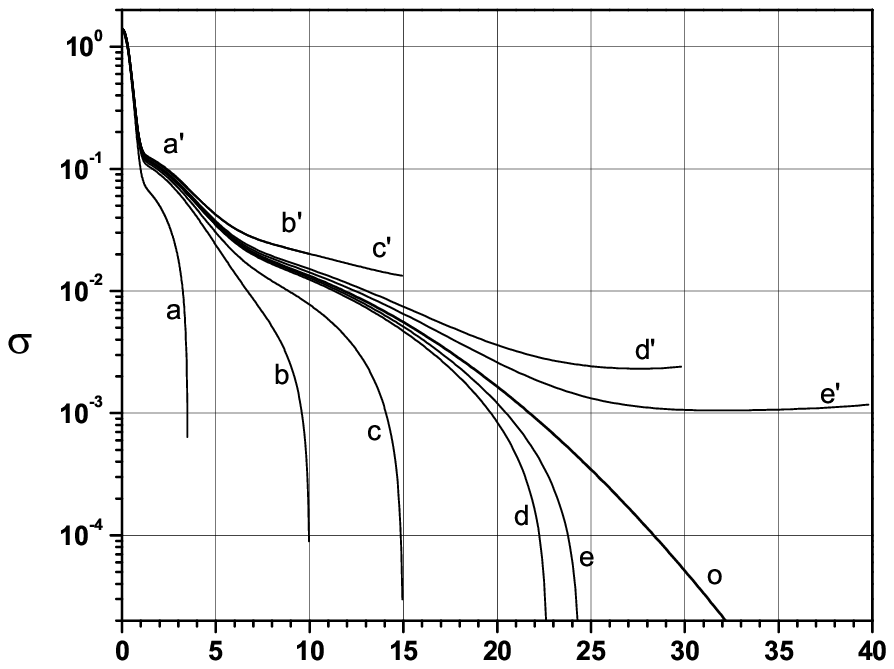}&
\includegraphics[width=0.333\textwidth,height=0.333\textwidth]{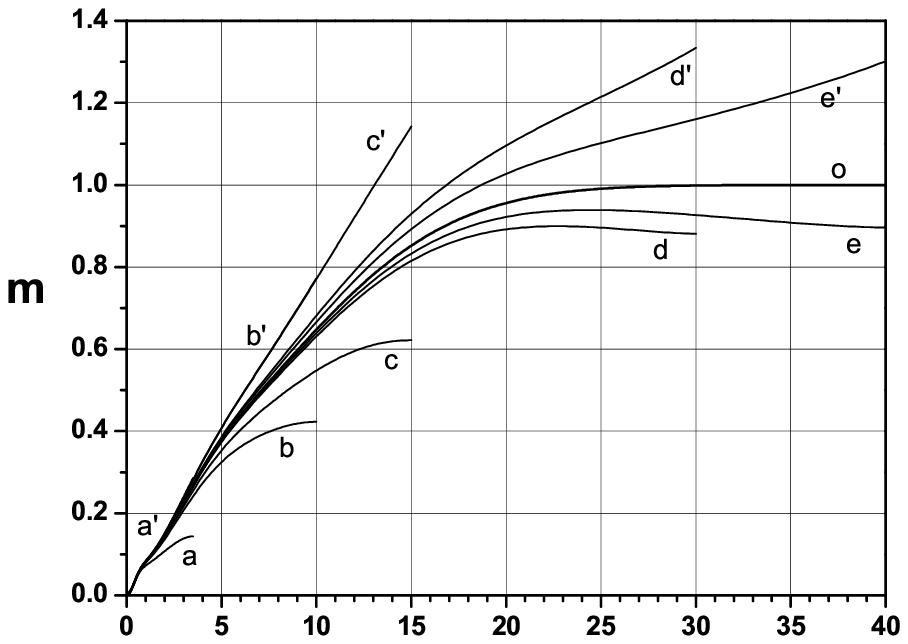}\\
I\,a)&I\,b)&I\,c)\\
\includegraphics[width=0.333\textwidth,height=0.333\textwidth]{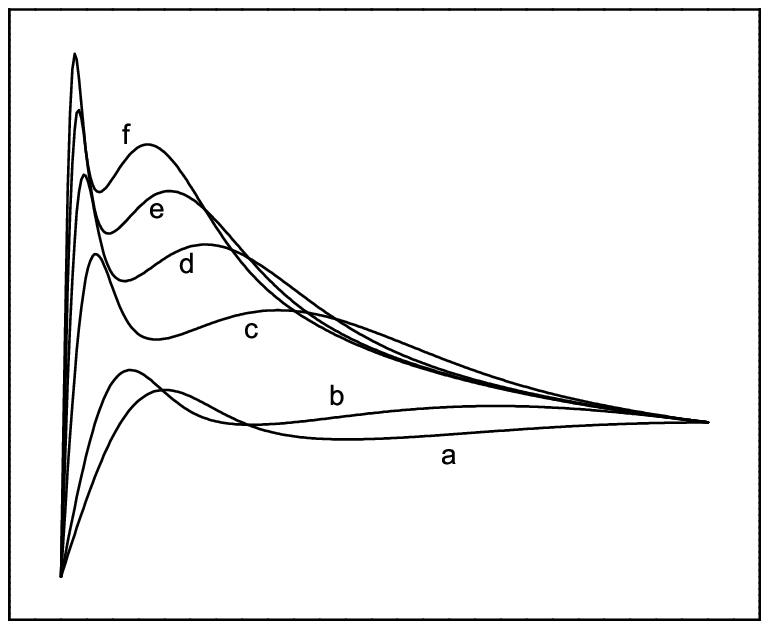}&
\includegraphics[width=0.333\textwidth,height=0.333\textwidth]{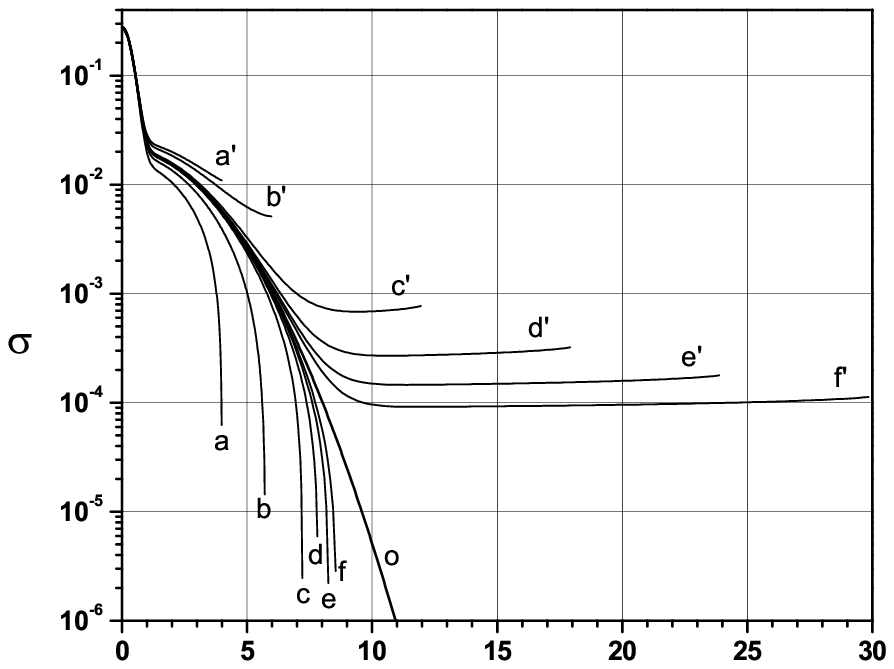}&
\includegraphics[width=0.333\textwidth,height=0.333\textwidth]{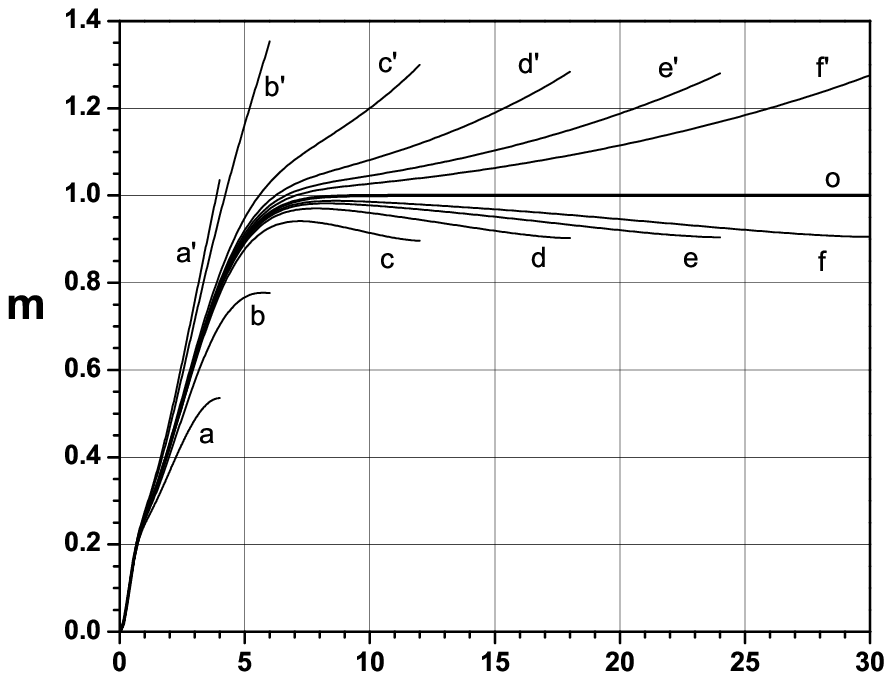}\\
  II\,a)&II\,b)&II\,c)\\
   \end{tabular}
 \caption{\label{fig:smpl_tri}The analysis of disk mass distributions which are
superpositions of: three gauss profiles discussed in figure
\ref{fig:trisigma30} (upper plots), and of two exponential profiles
(lower plots). The analysis illustrates the influence of the cutoff
radius on the accuracy of reconstruction of the original mass
distribution. The surface densities were calculated with the help of
integral (\ref{eq:sigmar}) (the lines marked by primed letters)
and with the help of the spectral representation (\ref{eq:sigmaseries}) of the rotation
curves (the lines without primes).
{\bf a)} Various pieces of the corresponding exact rotation curves
obtained by cutting off external parts at different radii (the
curves are shown in normalized variables $(x,u)$ explained in the
text). {\bf b)} The reconstructed surface mass densities and the
original density (denoted by 'o'). {\bf c)} The
corresponding mass functions, and the exact asymptotically unit
mass function.}
\end{figure*}

In figure \ref{fig:trisigma30} compared are different
approximations of the surface mass density obtained with the help of various
methods, for the rotation curve of a model disk with a known mass
distribution.
To illustrate the influence of the spatial extension of the rotation curve measurements
on the accuracy of the reconstruction of the mass distribution, in figure
\ref{fig:smpl_tri} presented is a similar analysis for various
samples of the same model rotation curve that was cut off at
different radii.
As follows from the resulting mass functions,
the cutoff error can be relatively large even in a distant empty
region, where the rotation curve is not yet Keplerian, that is,
when the cutoff error can not be estimated yet and taken into
account. For example, in the surface density diagram in
figure \ref{fig:trisigma30}, the line representing
$\widetilde{\sigma}_R(\rho)$ corrected for the cutoff error
(\ref{eq:deltasigma}), does not overlap with the exact value of
the surface density. Even though $99,9\%$ of the total mass is
concentrated in the disk of radius $R=30$, the rotation curve is still
not Keplerian outside this radius. At this radius equation
(\ref{eq:sigRbar}) overestimates the true mass by $33\%$, whereas
another estimation based on the spectral approximation given in
appendix \ref{app:spectral}
underestimates the true mass by $12\%$ only.

\section{Mass--to--light ratio and the results for NGC
5457} The mass distribution of the luminous matter in different galaxy
subsystems is customarily assumed to be proportional to the
corresponding local brightness via constant mass-to-light ratios
(frequently the blue band is chosen). This arbitrary assumption
ignores the dynamical information about the mass distribution encoded in
the non-monotonic features of rotation curves. This in turn, may result in wrong
distribution of matter among various
galactic components. Especially important are the distant regions where a massive
spherical hallo is always assumed, whereas in some cases a disk component of
baryonic matter may be still more pronounced. The best illustration of this is again the galaxy NGC 4736 \citep{bib:bratek}. Therefore, it seems better not to correlate the $M/L$ ratio with the mass distribution and derive it as a
point-dependent function from the surface density determined in
another way. Then one would be able to trace the presence of different
star populations as well as the dark matter distribution.

\begin{figure*}
\begin{tabular}{|@{}c@{}|@{}c@{}|}
\includegraphics[width=0.4\textwidth]{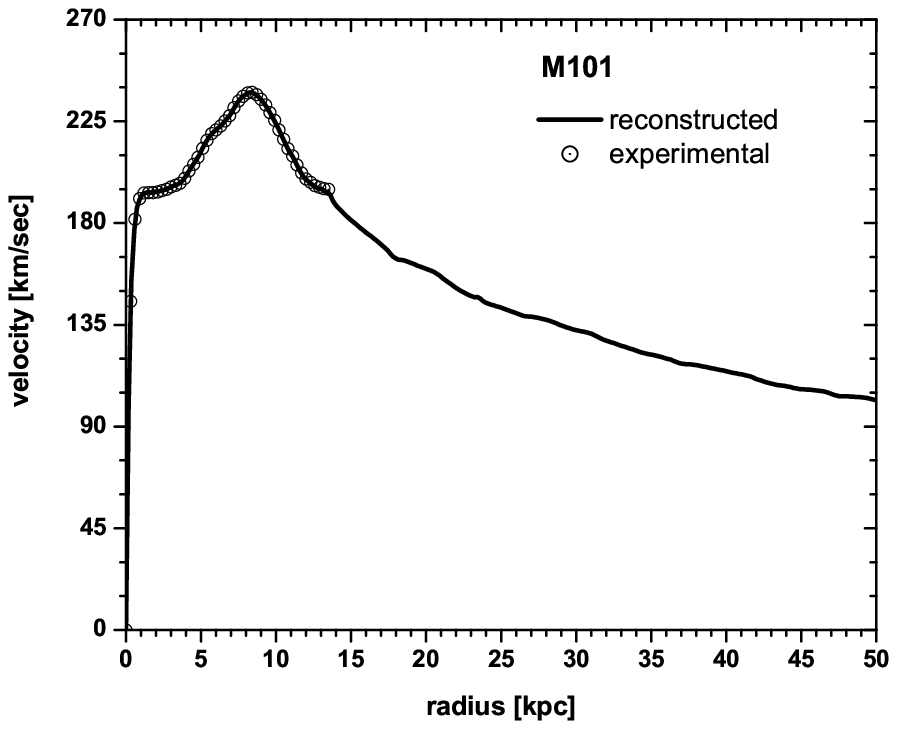}&
\includegraphics[width=0.4\textwidth]{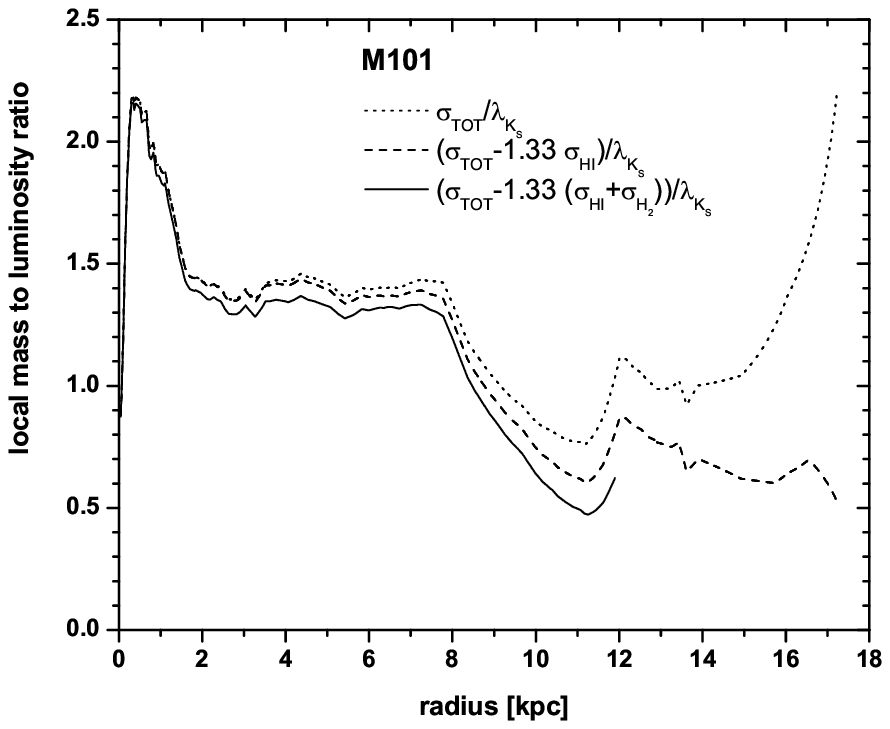}\\
a)&d)\\
\includegraphics[width=0.4\textwidth]{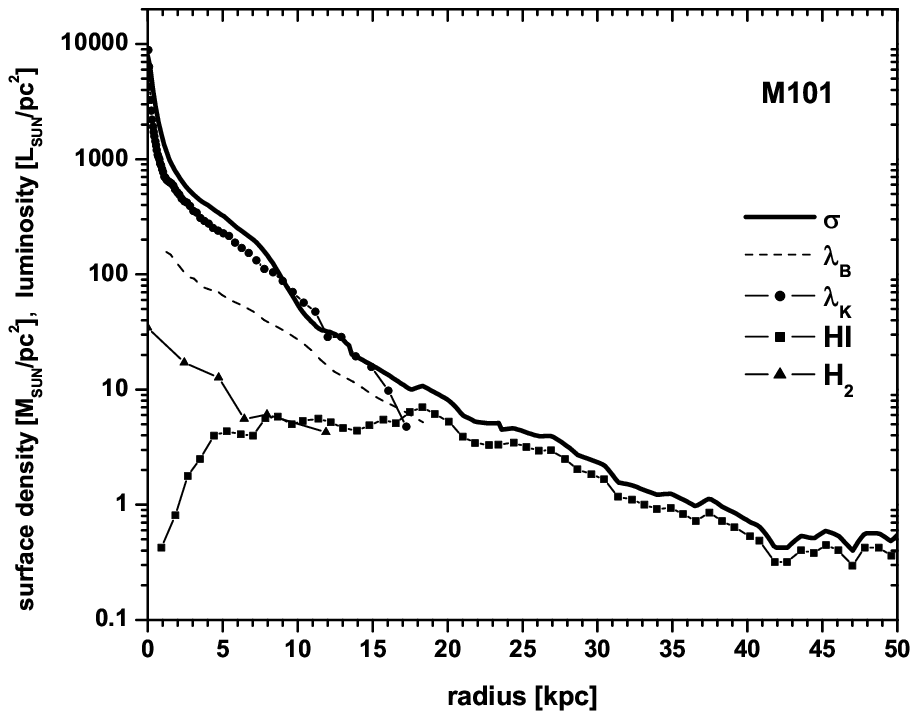}&
\includegraphics[width=0.4\textwidth]{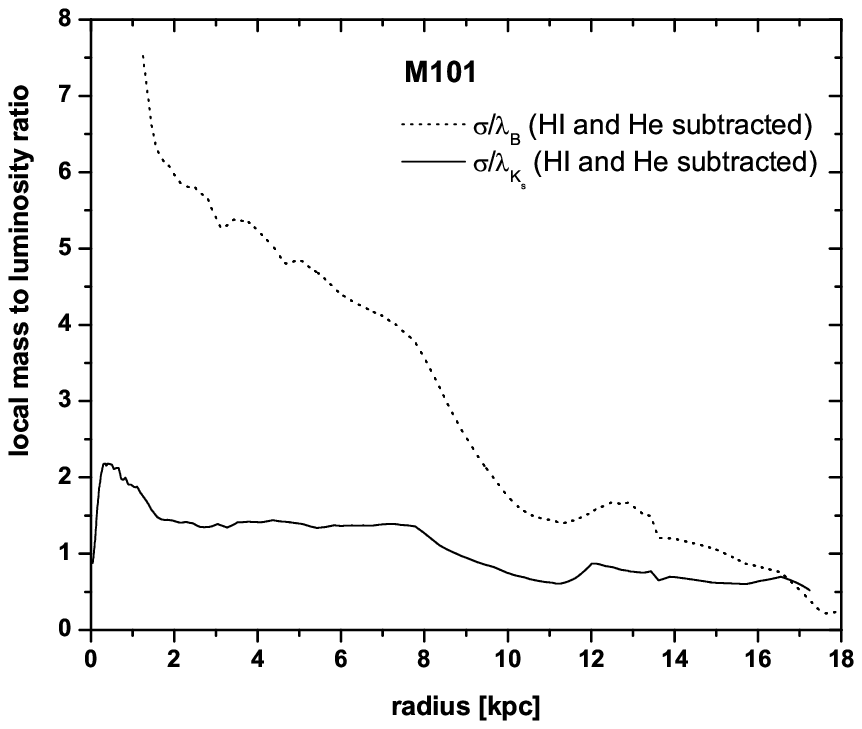}\\
b)&e)\\
\includegraphics[width=0.4\textwidth]{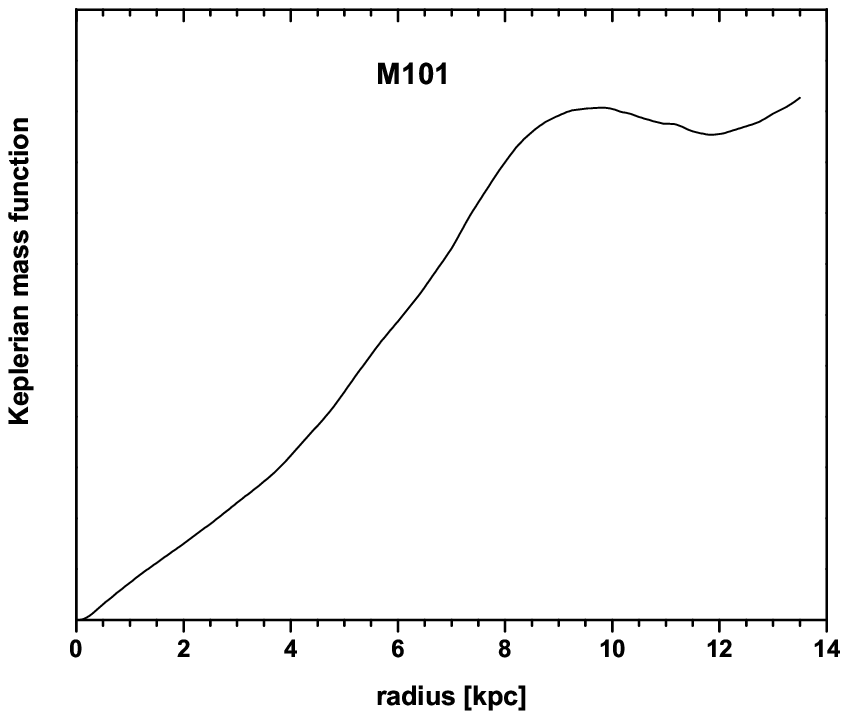}&
\includegraphics[width=0.4\textwidth]{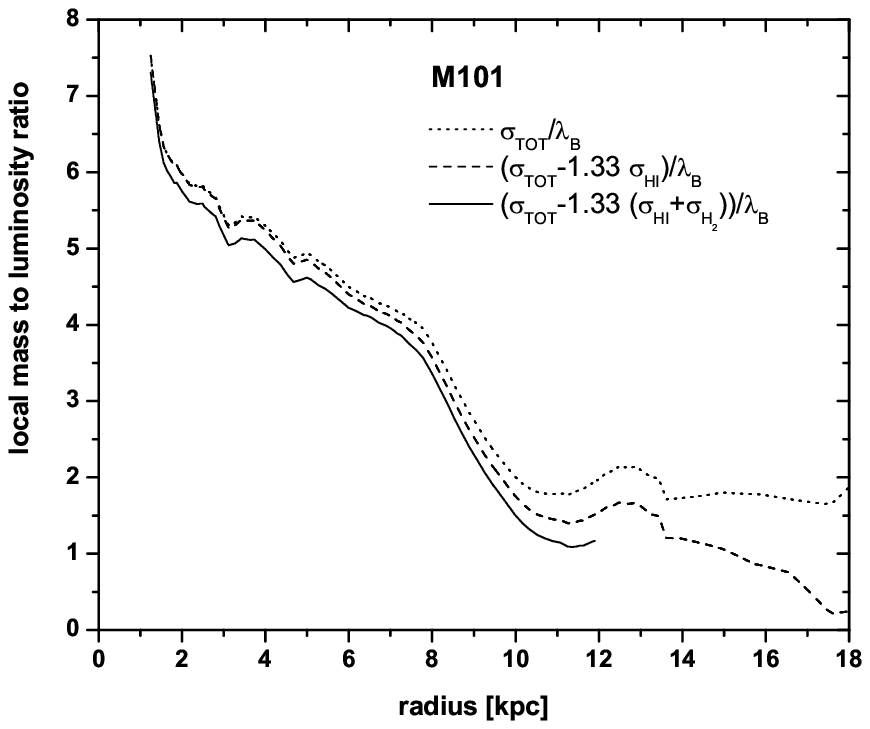}\\
c)&f)\\
\end{tabular}
\caption{\label{fig:m101} The results for spiral galaxy NGC 5457 (M101) obtained in the thin disk model  with the use of the iterative method developed in
\citep{bib:bratek}. \textbf{a)} the rotation curve calculated from
equation (\ref{eq:vfromsig}) for a surface mass density marked by
$\sigma$ in figure b); \textbf{b)} the global surface mass density
found by applying the iteration method, the hydrogen surface densities
for $HI$, $H_2$, and the surface luminosity profiles ($\lambda$) (K
and B bands); \textbf{c)} the Keplerian mass function
(\ref{eq:criterion}) for the observed part of rotation curve of NGC 5457;
\textbf{d)} the mass-to-luminosity ratio profile ($\sigma/\lambda$) in
$K_s$ filter for total surface density and with  hydrogen $HI$,
$H_2$ and helium subtracted (since they are not seen in the $K_s$ band);
\textbf{e)} comparison of local mass-to-luminosity profiles
($\sigma/\lambda$) in $B$ and $K_s$ bands (hydrogen and helium
subtracted); \textbf{f)} the mass-to-luminosity ratio profile in $B$
filter for the total surface density and the other one with hydrogen
$HI$, $H_2$ and helium subtracted}
\end{figure*}

NGC 5457 (Hubble type Sc I) is one of the largest nearby spiral galaxies, thus
also well studied \citep{bib:m101bosma}. The behavior of the
local $M/L$ ratio for this galaxy inferred from the $K_s$ band differs from
that in the $B$ band, c.f. figure \ref{fig:m101}d.\footnote{These
results were reported in part at the XLVII Cracow School of
Theoretical Physics in Zakopane
\citep{bib:zakopane}.}$^{,}$\footnote{The measurement data for NGC
5457 were taken from: the $HI$ surface density \& the $L_B$ luminosity
profile \citep{bib:lb}, the galaxy distance \citep{bib:dist}, the $H_2$
surface density \citep{bib:h2}, and the
 rotation curve \citep{bib:sof_www}.}
  The reconstruction of the surface mass density strongly depends on the chosen luminosity band -- in the most part of NGC 5457, the local $M/L$ ratio
calculated from the $B$ band is significantly greater than that from the $K_s$ band. What's more, low-mass, lower temperature stars make up the most of galactic disk.
Therefore, the global $M/L$ ratio in the
infrared band, like I, K, i, etc, is the best
indicator of the presence of dark matter in spiral galaxies, with
$M/L < 2$ being typical for normal stars.

Although the galaxy NGC 5457 is very large, its rotation curve has
been determined only out to 14 kpc (NGC 5457  has an asymmetric
Doppler image). The rotation curve breaks the sphericity condition
(\ref{eq:sph_cond}), c.f. figure \ref{fig:m101}, thus, for the
reasons discussed in  the previous sections, we use the global
disk model. To obtain the global mass distribution in this galaxy
we applied the method presented in
\citep{bib:bratek} that combines through iterations the measured rotation curve together
with gas measurements far from the galactic center.
The method gives
a self-consistent global mass distribution in the disk. Self-consistency means that a) the rotation curve calculated from the global surface mass density with the help of integral
(\ref{eq:vfromsig}) overlaps with the measured rotation curve, b) the calculated global surface mass density outside the cutoff radius overlaps with the amount of hydrogen and helium visible in the outer parts of this galaxy and c) the calculated global rotation curve and global surface density are mutual transforms of each other according to integrals (\ref{eq:vfromsig}) and (\ref{eq:sigmamoja}).

The global
characteristics of the analysis for NGC 5457 are given in table \ref{tab:a}.
\begin{table*}
\centering
\begin{tabular}{|c|c|}
\hline
$M_{TOT}$&$1.13\times10^{11}M_{\sun}$\\
$M_{HI}$&$1.5\times10^{10}M_{\sun}$\\
$M_{H_2}$&$3.4\times10^{9}M_{\sun}$\\
$L_{K_S}$&$7.45\times10^{10}L_{\sun}$\\
$L_{B}$&$2.49\times10^{10}L_{\sun}$\\
$M_{TOT}/L_{K_S}$&1.52\\
$M_{TOT}/L_{B}$&4.54\\
$M_{STAR}/L_{K_S}$&1.27\\
$M_{STAR}/L_{B}$&3.8\\
\hline
\end{tabular}
\caption{\label{tab:a}The other results for NGC 5457. $M_{K_S\sun}=3.45$,
$M_{STAR}=M_{TOT}-\frac{4}{3}\br{M_{HI}+M_{H_2}}$,
$D=7.2\mathrm{Mpc}$, inclination $18\mathrm{deg}$}
\end{table*}
In combination with the observed local brightness $\lambda$, the
determined surface density $\sigma$ yields the corresponding local
mass-to-light ratio profile $\sigma/\lambda$.  In the $K_S$ band,
$\sigma/\lambda$ increases near the disk edge, c.f. figure
\ref{fig:m101}. A similar increase was observed by Sofue for
several other galaxies \citep{bib:TakamiyaSofue}. This rise would
seem at first as the signature of the presence of dark matter in
the outer part of the galaxy. Indeed, if stars were to comprise the whole
baryonic mass of the galaxy, the $\sigma/\lambda$ ratio would be
always that of stars, even in the limit  $\lambda \to 0$. This is
not the case for NGC 5457, however. The increase of the local
mass-to-light ratio in this galaxy is caused by the presence of
large amounts of hydrogen beyond the stellar disk. If there is
some gas, eg. hydrogen and helium, then the local mass--to--light
ratio profile
$(\sigma_{stars}+\sigma_{\mathrm{H},\mathrm{He}})/\lambda$
diverges as $ \lambda \to 0$. But the gas contribution should be
subtracted as it is nonluminous in the visible and infrared bands.
As is seen in figure \ref{fig:m101}, after subtraction of the hydrogen
and primordial helium contribution, the local mass--to--light ratio
profile stops to increase at the disk edge. Also  in the $B$ band
the ratio has decreased. This fact, together with the low mass--to--light ratio in the $K_s$ band, shows that the $CDM$ halo is
not needed to account for this galaxy's rotation. Furthermore,
the fact of breaking the sphericity condition, practically
excludes a massive and spherical halo of CDM in NGC 5457.

\section{Concluding remarks}
For the accuracy of determination of the mass distribution and its
features in the disk component, local non-monotonic
properties of rotation curves are important. They are not less important than
the value of the rotational velocity. In particular, a local increase in
the velocity field may be caused simply by a local decrease of
matter density in the disk component and not by the increase of
density in the spherical component of a galaxy. The important
information about mass distribution carried by rotation curves may
be simply overlooked if one assumes that the local mass distribution
is proportional to the local brightness. In some cases the dark
matter halo is introduced simply because the amount of luminous
matter with the constant mass-to-light ratio found by the least
square fitting method, cannot account for the rotation at large
radii. The example of the spiral galaxy NGC 4736 discussed in
\citep{bib:bratek} very well illustrates this observation. This
shows also that determination  of mass distribution in spiral
galaxies is model dependent.  For more reliable predictions it is
thus important to have a generic model, more flexible and general
than the parametric models with smooth profiles and constant
mass-to-light ratios. It seems that the global thin disk model
provides such a generic model, at least for flattened galaxies with
rotation curves breaking at large radii the sphericity condition.

\medskip
Due to observational cutoff in rotation data, the surface mass density
reconstruction in the galactic disk (and consequently in the whole
galaxy), is subject to high uncertainties, which increase with the distance from the center.
Only in particular situations of rotation curves with almost
Keplerian tails, the cutoff errors can be eliminated and taken
into account. The observationally determined part of a given
rotation curve may be explained by various internal disk mass
distributions depending on how the rotation law has been
extrapolated beyond the cutoff radius. This uncertainty can be
removed if additional constraints on the mass distribution in the
outer parts, not covered with rotation measurements, are taken into
account. In this respect the hydrogen distribution in  the outer
regions can be used. We have established for galaxies NGC4736 and
NGC5457 that the observed amount of baryonic matter accounts for rotation of these galaxies in the approximation of the global disk model.


\appendix

\section{Basic equations of the axisymmetric thin disk model}
\label{app:deriv} We assume an axisymmetric mass distribution over an
infinitely thin disk. The cylindrical coordinates $(\rho,\phi,z)$ are
chosen such that the disk overlaps with the $z=0$ plane.
Gravitational potential $\Phi$ is $z$-symmetric, hence it suffices
to consider the half space $z>0$. In this region the complete
function space of axisymmetric, everywhere bounded and vanishing
at infinity functions is spanned by the base solutions
$J_0(\omega\rho/L)\exp{-\omega{z/L}}$ of the Laplace equation,
with $\omega>0$ and $L$ being a length scale. Linearity implies
that the most general solution in this function space can be
represented as a superposition
\begin{equation} \label{eq:phiintegral}
\Phi(\rho,z)=-2\pi{}v_L^2\int\limits_0^{\infty}\hat{\sigma}(\omega)
J_0\br{\omega\frac{\rho}{L}}
\mathrm{exp}\br{-\omega\frac{z}{L}}\ud{\omega},
\end{equation}
$v_L$ is a velocity scale, and the spectral amplitude
$\hat{\sigma}$ is dimensionless. A dimensionless distance $x$ and a
velocity function $u(x)$ are defined by \[x=\frac{\rho}{L},\qquad
u(x)=\frac{v(Lx)}{v_L}.\] The discontinuity of $\partial_z\Phi$ at
$z=0$ is interpreted as an infinitely thin layer of mass spread over
the disk's plane with the surface mass density
\[\sigma(\rho)=\frac{1}{2\pi{}G}\partial _{z}\Phi |_{z=0},\] hence
\begin{equation}\label{eq:sigmaintegral}
\sigma(\rho)= \frac{v_L^2}{GL
}\cdot\int\limits_0^{\infty}\omega\hat{\sigma}(\omega)
J_0\br{\omega\frac{\rho}{L}}\ud{\omega}.
\end{equation}
Note, that $\hat{\sigma}(\omega)$ is the spectral amplitude of
$\sigma(\rho)$ (the inverse Hankel transform)
\begin{equation}\label{eq:sigampl}
\hat{\sigma}(\omega)=\frac{GL}{v_L^2}\int\limits_0^{\infty}
x\sigma(Lx)J_0(\omega{x})\ud{x}.
\end{equation}
The condition for coplanar, circular and concentric orbits in the
$z=0$ plane reads $v^2(\rho)=\rho\,\partial_{\rho}\Phi(\rho,0)$
for all $\rho$. Hence, having formally differentiated
(\ref{eq:phiintegral}), one gets for $z=0$
\begin{equation}\label{eq:ufromsigampl}
u^2(x)=2\pi{x}\int\limits_0^{\infty}\omega
\hat{\sigma}(\omega)J_1(\omega{x})\ud{\omega},
\end{equation} and the resulting inverse relation
\begin{equation}
\label{eq:sigmaomega} \hat{\sigma}(\omega)=
\frac{1}{2\pi}\int\limits_0^{\infty}u^2(x) J_1(\omega{x})\ud{x}.
\end{equation}
On substituting (\ref{eq:sigmaomega}) to (\ref{eq:sigmaintegral}), integrating by parts,
provided that $v(\rho)$ is differentiable, that
$v^2(\rho)\to0$ as
$\rho\to0$, that $v^2(\rho)/\sqrt{\rho}\to0$ as
 $\rho\to\infty$,\footnote{These assumptions assure at the same time
 the existence of integral
(\ref{eq:sigmaomega}) and the suitable boundary conditions required for
the equivalence of integrals (\ref{eq:toomre_app}) and
(\ref{eq:sigmamoja_app}) (flat rotation curves are incompatible with these conditions). To assure at the same time mutual invertibility of integrals (\ref{eq:ufromsigampl}) and (\ref{eq:sigmaomega}) one needs a stronger condition that $\lim\limits_{\rho\to\infty}\sqrt{\rho}\,v^2(\rho)\to0$. } and using the upper integral in
footnote\footnote{ \label{ft:tab}
\[\begin{array}{l}
\int\limits_0^{\infty}\ud{\omega}J_0(\omega{x})
J_0(\omega\xi)=\left\{\begin{array}{ll}
\frac{2}{\pi\xi}K\br{\frac{x}{\xi}},& 0<x<\xi\\
\frac{2}{\pi{}x}K\br{\frac{\xi}{x}},& 0<\xi<x\\
\end{array}\right.\\
\int\limits_0^{\infty}\ud{\omega}J_1(\omega{\xi})
J_1(\omega{x})=\left\{\begin{array}{ll}
\frac{2}{\pi\xi}\br{K\br{\frac{\xi}{x}}-E\br{\frac{\xi}{x}}       },& 0<\xi<x\\
\frac{2}{\pi{x}}\br{K\br{\frac{x}{\xi}}-E\br{\frac{x}{\xi}} },&
0<x<\xi,
\end{array}\right.
\end{array}\]
cf. \citep{bib:15}.}, we arrive at the Toomre integral
\begin{eqnarray}\label{eq:toomre_app}
{\sigma}(\rho)=\frac{1}{\pi^2G}\vp \left[
\int\limits_0^{\rho}\frac{\ud{v^2}(\chi)}{\ud{\chi}}
\cdot\frac{1}{\rho}K\br{\frac{\chi}{\rho}}\ud{\chi}
\right.\nonumber\\\left.+\int\limits_{\rho}^{\infty}\frac{\ud{v^2}(\chi)}{\ud{\chi}}
\cdot\frac{1}{\chi}K\br{\frac{\rho}{\chi}}\ud{\chi}\right].
\end{eqnarray}
The integral should be understood in the principal value sense
(the symbol $\vp$ indicates this kind of integration), thus the singularity of the elliptic function $K(y)$ at
$y=1$ does not contribute. The equivalent form of the Toomre
integral (with the same assumptions about $v(\rho)$ as above) reads
\begin{eqnarray}\label{eq:sigmamoja_app}
\sigma(\rho)=\frac{1}{\pi^2G} \vp \left[\int\limits_0^\rho
v^2(\chi)\biggl(\frac{K\br{\frac{\chi}{\rho}}}{ \rho\
\chi}-\frac{\rho}{\chi}
\frac{E\br{\frac{\chi}{\rho}}}{\rho^2-\chi^2}\biggr)\ud{\chi}
\right.\nonumber\\ +\left.\int\limits_\rho^{\infty}v^2(\chi)
\frac{E\br{\frac{\rho}{\chi}}
}{\chi^2-\rho^2}\,\ud{\chi}\right].\end{eqnarray}
The integrand has an ordinary pole
$\frac{v^2(\rho)}{2\rho(\chi-\rho)}$ at $\chi=\rho$, which is
easily integrable numerically and dominates the singularity of the
elliptic function $K$. The above equation can be derived analogously like the Toomre integral, however it can be also proved by integration by parts the
integral (\ref{eq:toomre_app})  (in
order to use the relevant theorems of integral calculus, the
integral (\ref{eq:toomre_app}) should be first rewritten as
$\int_{0}^{x-\epsilon}+\int_{x+\epsilon}^{\infty}$, then
integrated by parts, and finally the limit $\eps\to0$ should be
taken).

The inverse of integral (\ref{eq:sigmamoja_app}) reads
\begin{eqnarray}\label{eq:vfromsig_app}
\frac{v^2(\rho)}{4\,G\,\rho}=\vp\left[
\int\limits_{0}^{\rho}\sigma(\chi)\frac{\chi{}E
\br{\frac{\chi}{\rho}}}{\rho^2-\chi^2}\ud{\chi}\right.
\nonumber\\-\left. \int\limits_{\rho}^{\infty}
\sigma(\chi)\biggl(
\frac{\chi^2E\br{\frac{\rho}{\chi}}}{\rho\br{\chi^2-\rho^2}} -
\frac{K\br{\frac{\rho}{\chi}}}{\rho}\biggr)\ud{\chi}\right].
\end{eqnarray}
Indeed, by substituting (\ref{eq:sigampl}) in
(\ref{eq:ufromsigampl}) one gets
\[\frac{u^2(x)}{x}=2\pi{}\int\limits_0^{\infty}
 J_1(\omega{}x)\ud{\omega}
 \br{\frac{GL}{v_L^2}\cdot
\int\limits_0^{\infty}
\sigma(Ly)\,\partial_y\br{yJ_1(\omega{}y)}\ud{y}}\] where the
identity
$\omega^{-1}\partial_y\br{yJ_1(\omega{y})}=yJ_0(\omega{y})$ has
been used. Next, on integrating by parts, with the assumption that
$\rho^2\sigma(\rho)\to0$ as $\rho\to0$ and
$\sqrt{\rho}\sigma(\rho)\to0$ as $\rho\to\infty$, we obtain
\[\frac{u^2(x)}{x}=\frac{GL}{v_L^2}\br{-2\pi\int\limits_0^{\infty}
\xi\ud{\xi}\partial_{\xi}\sigma(L\xi)\int\limits_0^{\infty}
J_1(\omega\xi)J_1(\omega{}x)\ud{\omega}}.\] The rightmost integral is given in
footnote$^{\ref{ft:tab}}$. On integrating
by parts again, with the same assumptions about the limiting behavior of
$\sigma(\rho)$ as before, one arrives at the final result
(\ref{eq:vfromsig_app}). When doing these calculations, the same
integration splitting and taking the limit, as in the derivation of
(\ref{eq:sigmamoja_app}), applies.

A simple analytic example using the above formulas and illustrating the equivalence of (\ref{eq:toomre_app}) and
(\ref{eq:sigmamoja_app}) was given in section \ref{sec:example}.

\section{A discrete spectral representation}\label{app:spectral}
For the observational reasons, a rotation curve $v(\rho)$ of a galaxy is known
only for radii $\rho\in(0,R)$, $R$ is the data cutoff radius. Let's
define $x=\rho/R$, $u(x)=v(Rx)/v_R$ with $v_R\equiv{}v(R)$. The
function $u^2(x)/x$ is thus defined on the unit interval
$x\in(0,1)$ and $u(1)=1$. Under some assumptions the function $u^2(x)/x$ can
be represented in this interval as a series of a complete set of orthogonal
functions on this interval. Although we may choose any complete set of functions
satisfying the appropriate boundary conditions on this interval, best
suited for the disk symmetry are cylindrical functions. If
$u^2(x)/\sqrt{x}$ is integrable in the interval $x\in(0,1)$ then
$u^2(x)/x$ can be represented in this interval as a series of
Bessel functions \citep{bib:lebiediev}. To
conform with the more general integral representation
(\ref{eq:ufromsigampl}) we choose the Bessel function of the first
order, then
\begin{equation} \label{eq:vseries}
\frac{u^2(x)}{x}=\sum_{k}\check{\sigma}_k\,J_1\br{\omega_k\,x},\quad
0<x<1,\quad J_0(\omega_k)=0.
\end{equation}
The summation is taken over all positive zeros $\omega_k$ of the
Bessel function of the zeroth order $J_0(x)$. On multiplying both
sides by $x\,J_1(\omega_m\,x)$ and integrating over $x\in(0,1)$
one gets
\begin{equation}
\label{eq:expanscoeff}
\check{\sigma}_k=\frac{\omega_k}{J_1(\omega_k)}\,\mu_k,\qquad
\mu_k=\frac{2}{\omega_k}\int\limits_0^1u^2(x)
\frac{J_1(\omega_k\,x)}{J_1(\omega_k)}\,\ud{x}.\end{equation} The
countable set of $\check{\sigma}_k$'s suffices to describe $v(\rho)$
completely only for $\rho<R$. However, this information is
insufficient to obtain the continuous spectrum
$\hat\sigma(\omega)$ of the true global $\sigma(\rho)$, thus even
for $\rho<R$ one is unable to determine $\sigma(\rho)$ from the
observed part of rotation curve. This can be done only
approximately. For example, assuming that $v(\rho)=0$ for
$\rho>R$, we obtain from (\ref{eq:sigmaomega})
\[\hat{\sigma}(\omega)=\frac{1}{2\pi}\sum\limits_k\check{\sigma}_k\frac{\omega
{}J_0(\omega)J_1(\omega_k)}{\omega_k^2-\omega^2}.\] Since
$J_0(\omega)\approx{}J_1(\omega_k)(\omega_k-\omega)$ when
$\omega\approx\omega_k$, the spectrum is continuous. Another
extension of $v(\rho)$ beyond $R$ arises by assuming that
expansion (\ref{eq:vseries}) is formally valid also for $\rho>R$,
then we obtain from (\ref{eq:sigmaomega}) a discrete spectrum
\begin{equation}\label{eq:discretespect}
\hat{\sigma}(\omega)=\frac{1}{2\pi}\sum\limits_k\check{\sigma}_k\frac{
\delta(\omega-\omega_k)}{\omega}.\end{equation} It should be
clear, that in a similar way, one can construct from the
$\check{\sigma}_k$'s set, infinitely many other spectra
$\hat{\sigma}(\omega)$. They will correspond to different global surface
densities giving the same rotation law for $\rho<R$ and different
for $\rho>R$. In particular, this is the case for the global
surface mass densities obtained by taking the inverse transform of
the above two spectra.\footnote{The fact that a function defined
on the whole real axis, when restricted to a finite interval can
be represented by many discrete and continuum spectra, should not
astonish. To give a simple example, consider a function such that
$f(x)=\sin{x}$ for $x\in(0,\pi)$ and $f(x)=0$ elsewhere. Then
\[\hat{f}(\omega)=\frac{1+\exp{-i\pi\omega}}{2\pi(1-
\omega^2)}\qquad \mathrm{and}\qquad\hat{f}(\omega)=\frac{1}{i} \delta(\omega^2-1)\]
are distinct spectral representations of this function on the
interval $(0,\pi)$, that is, the inverse transform
$\int_{-\infty}^{+\infty}\hat{f}(\omega)
\exp{i\omega{}x}\ud{\omega}$ converges to $\sin(x)$ in this
interval for both the spectra. Note, however, that outside the
interval the integral converges to different functions.} However, the closer to the galactic center the less the
two mass distributions differ from each other. Indeed, the error
of determining $\sigma$, which is calculated from equation
(\ref{eq:deltasigRbar}) in the limit of small radii, reads
\[\lim\limits_{\rho\to0}\Delta\widetilde{\sigma}_R(\rho)=\frac{v_R^2}{RG}\cdot\frac{1}{2\pi}\int\limits_{1}^{\infty}\ud{\xi}
\frac{u^2(\xi)-1}{\xi^2}, \qquad u(1)=1,\] thus should be much
smaller than the characteristic density scale $v_R^2/(RG)$ (as was
shown in section \ref{sec:cutofef} it is about $8\%$ of the scale
for Keplerian falloff), while the central density is usually much
greater than this scale.

With this $\hat{\sigma}(\omega)$ we obtain from (\ref{eq:sigmaintegral}) the following series
\begin{equation}\label{eq:sigmaseries} \Sigma(\rho)=
\frac{v^2_R}{GR}\cdot\frac{1}{2\pi}\sum\limits_k\check{\sigma}_kJ_0\br{
\omega_kx}, \qquad x=\frac{\rho}{R}<1,
\end{equation} where $\check{\sigma}_k$'s have been calculated with
the help of (\ref{eq:expanscoeff}). Let $\rho_1>0$ be the smallest
radius  at which $\Sigma(\rho)$ becomes negative. By definition one
sets in this approximation of the true $\sigma(\rho)$ that $\sigma(\rho)\approx\sigma_A(\rho)$ where $\sigma_A(\rho)=0$ for
$\rho>\min(\rho_1,R)$, and $\sigma_A(\rho)=\Sigma(\rho)$ for
$0<\rho<\min(\rho_1,R)$. Then close to the center, the integral
(\ref{eq:vfromsig_app}) with $\sigma(\rho)=\sigma_A(\rho)$, quite
well approximates the original rotation curve that was cut off at
$R$ (c.f. section \ref{sec:cutofef} for more details).

\end{document}